\documentclass{article}
\usepackage{geometry}
\usepackage{amsmath} 
\usepackage{amssymb} 
\usepackage{amsfonts} 
\usepackage{mathtools} 
\usepackage{hyperref}
\usepackage{caption}
\usepackage{subcaption}
\usepackage{graphicx}
\usepackage{tikz}
\usetikzlibrary{angles,quotes}
\usepackage{float}
\usepackage{environ} 
\usepackage{setspace}
\usepackage{framed}

\usepackage[most]{tcolorbox}

\usepackage{biblatex}
\usepackage{color}
\usepackage{xcolor}
\usepackage{soul}
\usepackage{enumerate}
\usepackage{enumitem}

\usepackage{algpseudocode}

\usepackage{lineno}

\bibliography{references.bib}

\title{Three-dimensional mesh adaptation in PFEM\\
\large Preprint}
\author{\textbf{Thomas Leyssens, Jonathan Lambrechts, Jean-François Remacle} \\
Institute of Mechanics, Materials and Civil Engineering, UCLouvain, Belgium \\
thomas.leyssens@uclouvain.be}
\date{December 23, 2025}

\begin{document}
\setstretch{1.125}

\maketitle

\begin{abstract}
    Chaotic free surface flows are challenging problems to simulate numerically, mainly due to the significant changes in geometry and frequent topological changes. 
Methods that track the evolution of the fluid in a Lagrangian formulation are a natural choice. 
One such method is the Particle Finite Element Method (PFEM). 
As a hybrid particle-based and mesh-based method, PFEM leverages advantages from both approaches. 
The equations of motion are solved on a mesh using the finite element method and the obtained velocity field is used to displace the nodes of this mesh, considered as particles carrying all the relevant information across time steps. 

To avoid element distortion, the mesh is frequently re-generated. 
This introduces some challenges: 
How can the new shape of the domain be detected?
How can the quality of the elements be kept acceptable? 
Can adaptive mesh refinement increase the  accuracy and efficiency of the solver?
Can PFEM simulations be performed in the presence of complex boundary geometries?

In this work, three contributions to the geometry and mesh component of PFEM are introduced for three-dimensional free surface flow simulations. 
First, we propose a different domain reconstruction approach than the classically used $\alpha$-shape procedure, namely through the use of the advected boundary from the previous time step as a predicate to represent the new shape of the domain. 
Second, an adaptive refinement procedure is proposed in two steps: refinement of the boundary surface followed by quality-based node insertion in the bulk.
Third, an approach for managing boundaries in complex geometries is presented. 
A series of applications is shown to demonstrate the interest of the approach. 

\end{abstract}

\section{Introduction}

The particle finite element method, or PFEM, is an interesting approach for free surface flow simulations.
Defined through a Lagrangian finite element framework coupled to an automatic surface tracking algorithm, the method allows to simulate strong topological changes of the simulated domain.
These features make it a method of choice for simulations such as fluid-structure interactions \cite{cerquagliaFSI}, landslide modelling \cite{vajont}, and melt-pool dynamics \cite{fevrier}. 

In PFEM, the domain evolution is described by Lagrangian particles, while the governing equations are discretized using the finite element method, benefiting from its mature and well-understood mathematical foundations.
In contrast to methods such as the material point method (MPM) \cite{mpm} or the particle-in-cell method \cite{pic}, which rely on a fixed background mesh over which particles are advected, PFEM operates on a mesh constructed from the Lagrangian particles themselves. The particles become the nodes of the finite element mesh, and the equations of motion are solved on this mesh using a classical piecewise linear finite element discretization.
The positions of the particles are subsequently updated by integrating the computed velocity field in time.

Keeping the same set of particles throughout the simulation is not feasible, since even in incompressible flows, a shearing motion will progressively transform an initially well-organized particle distribution into a highly distorted and unusable configuration.
As a result, the particle distribution must be updated whenever necessary.
In the present work, this update is performed at every time step. 
Instead of operating directly on the particle cloud, PFEM performs the particle update by acting on the mesh obtained from the Delaunay triangulation of the particles, which is considerably more robust.

Repeated mesh regeneration endows PFEM with considerable flexibility, particularly through its ability to naturally accommodate topological changes of the fluid domain, but it also represents a major computational challenge.
In practice, the computational cost of triangulating a point cloud, even with several million points, is negligible compared to that of solving the incompressible Navier–Stokes equations on the resulting mesh \cite{marot2019one}. The difficulty lies instead in the fact that a Delaunay triangulation fills the convex hull of the point cloud, and this convex hull provides a particularly poor oracle for determining the actual shape of a fluid domain represented by a set of particles.
Moreover, as discussed above, the mesh must subsequently be adapted in order to restore a well-balanced and harmonious particle distribution. Finally, the set of particles defining the fluid domain evolves within a bounded region, typically delimited by rigid walls. When a particle reaches such a rigid boundary, an appropriate boundary condition must be enforced.

It therefore becomes clear that a robust PFEM formulation must address three key geometrical challenges:
\begin{framed}
\begin{enumerate}
  \item an accurate and efficient oracle to determine the shape of the fluid domain,
  \item a mesh adaptation strategy to update the particle distribution,
  \item a general and flexible representation of the solid boundaries confining the fluid motion.
\end{enumerate}
\end{framed}




The objective of this work is to present three innovative approaches for tackling these challenges.

First, we propose a new oracle to represent the fluid domain after re-meshing. 
The classical $\alpha$-shape approach has many known limitations, resulting in mass errors \cite{franciMass}. 
While retaining the algorithmic efficiency and robustness of the Delaunay triangulation, we use the advected boundary from the previous time step as a predicate to define the updated shape of the domain. This boundary representation, in the form of a deformed triangulation, is not necessarily watertight; nevertheless, it can be used as a robust oracle to answer the fundamental question of whether a point $\boldsymbol{x}$ lies inside or outside the fluid domain.
To this end, we draw inspiration from recent work on fast winding numbers~\cite{windingNumber2}.

Second, we present a 3D adaptive mesh refinement technique for PFEM. 
Using this technique, the quality of the elements remains sufficiently high at all times, and local mesh adaptation allows to concentrate the degrees of freedom of the simulation in the regions of interest, for example, along the free surface. 
The approach consists of first refining the bounding surface elements, followed by refinement of the internal volume elements. 
This mesh adaptation procedure is based on a size field, allowing the user have control over the mesh.

Third, we propose a robust treatment of solid boundaries that fully decouples
the representation of the fluid domain from that of the solid geometry.
Solid walls are neither meshed nor topologically connected to the fluid domain.
Instead, all interactions with the solid boundaries rely exclusively on two
geometric queries.
The first query is a ray--solid intersection operator. Given a point
$\boldsymbol{x}$ and a direction $\boldsymbol{d}$, it returns the first
intersection of the ray $(\boldsymbol{x},\boldsymbol{d})$ with the solid
boundary, if such an intersection exists. This operator is used to identify
fluid faces that are in contact with solid walls.
The second query is an orthogonal projection operator, which maps a point
$\boldsymbol{x}$ onto its closest point on the solid boundary. This operator is
used to enforce boundary conditions and to reposition particles near walls.
Because the solid geometry is accessed only through these two queries, it may
be represented in a wide variety of forms, including surface meshes, triangle
soups, level-set representations, or distance fields. This makes the approach
robust with respect to imperfect or non-watertight solid geometries and
significantly simplifies the handling of fluid--solid interactions.


In the following sections, section \ref{sec:eqt} first presents the physical context. 
The core of the paper then follows in section \ref{sec:numer}, where we present the new fluid oracle (section \ref{sec:oracle}), the mesh adaptation algorithm (section \ref{sec:adapt}), and the boundary management approach (section \ref{sec:bnd}). 
Some validations are presented in section \ref{sec:valid}, and illustrative applications are shown in section \ref{sec:applic}.

\section{Governing equations}\label{sec:eqt}
We consider a Newtonian, incompressible fluid. 
The momentum and mass conservation equations can be written as follows:
\begin{linenomath}
    \begin{align}
        \rho \frac{D\mathbf u}{Dt} &= \nabla  \cdot \boldsymbol \sigma - \rho \mathbf g \label{eq:mom}\\
        \nabla \cdot \mathbf u &= 0.\label{eq:mass}
    \end{align}
\end{linenomath}
In this expression, $\mathbf{u}$ represents the fluid velocity~$[\mathrm{m\,s^{-1}}]$, $\rho$ the density~$[\mathrm{kg\,m^{-3}}]$, and $\mathbf{g}$ the gravitational acceleration~$[\mathrm{m\,s^{-2}}]$.
The Cauchy stress tensor $\boldsymbol{\sigma}~[\mathrm{Pa}]$ is given by
\begin{linenomath}
\begin{align}
  \boldsymbol{\sigma} = - p \mathbf{I} + \mu (\nabla \mathbf{u} + (\nabla \mathbf u)^T),
\end{align}
\end{linenomath}
where  $\mu~[\mathrm{Pa\,s}]$ is the dynamic viscosity and $p~[\mathrm{Pa}]$ the pressure.
In the purely Lagrangian setting of PFEM, the position $\mathbf x$ of the particles is updated by advection, 
\begin{linenomath}
    \begin{align}
        \frac{D\mathbf x}{Dt} = \mathbf u.\label{eq:adv}
    \end{align}
\end{linenomath}
The system is closed by a set of Neumann and Dirichlet boundary conditions in terms of stresses and displacements: 
\begin{linenomath}
    \begin{align}
        \boldsymbol{\sigma}(\mathbf x, t) \cdot \mathbf{\hat n} &= \bar{\mathbf t}(\mathbf x, t), \hspace{.5cm} \forall \mathbf x \in \partial \Omega_N\\
        \mathbf u(\mathbf x, t) &= \bar{\mathbf u}(\mathbf x, t), \hspace{.5cm} \forall \mathbf x \in \partial \Omega_D
    \end{align}
\end{linenomath}
    
Equations \ref{eq:mom} and \ref{eq:mass} are solved together, while the advection equation \ref{eq:adv} is solved by moving the particles. 
The weak form is obtained through the Galerkin approximation, defining a set of suitable vector test functions $\mathbf v \in \mathbf V$ and scalar test functions $ q \in Q$.
\begin{linenomath}
    \begin{align}
        \begin{split}
            \rho \int_\Omega \frac{D \mathbf u}{Dt}  \cdot \mathbf v = -\mu \int_\Omega \nabla \mathbf u \cdot \nabla \mathbf v + &\int_\Omega \nabla p \cdot \mathbf v + \\ \rho \int_\Omega \mathbf g \cdot \mathbf v + &\int_{\partial \Omega_N} \mathbf{\bar t}\cdot \mathbf v, \hspace{.2cm} \forall \mathbf v \in \mathbf V \label{eq:weak_mom}
        \end{split}
    \end{align}
\end{linenomath}
\begin{linenomath}
    \begin{align}
        \int_\Omega (\nabla \cdot \mathbf u) q - \int_\Omega \tau_{\text{pspg}} \boldsymbol{r}_u \cdot \nabla q = 0, \hspace{.2cm} \forall q \in Q.\label{eq:weak_mass}
    \end{align}  
\end{linenomath}

We consider linear shape functions for velocity and pressure, and both equations are solved in a single linear system.
Hence, to satisfy the Ladyzhenskaya-Babuska-Brezzi (LBB) condition, the second term in \ref{eq:weak_mass} corresponds to a Petrov-Galerkin pressure stabilisation (PSPG) \cite{pspg}.
Defining $\boldsymbol{r}_u$ as a residual term and ommitting the viscous part:
\begin{linenomath}
    \begin{align}
        \boldsymbol{r}_u = \rho \frac{D \mathbf u}{Dt} + \nabla p - \rho \mathbf g.
    \end{align}
\end{linenomath}
    $\tau_{\text{pspg}}$ is a coefficient which results from the Petrov-Galerkin approach:
\begin{linenomath}
    \begin{align}
        \tau_{\text{pspg}} = \frac{1}{\sqrt{\frac{4}{\Delta t ^2} + \frac{4 \mu}{\rho h^2}}},
    \end{align}
\end{linenomath}
where $h$ is an element mesh size. 
We note that this coefficient causes the stabilisation term to vanish as the mesh size and time step tend to zero. 
 
Time integration of these equations is performed with an implicit Euler scheme.
The code is implemented in the in-house, open-source solver Migflow, and further details of the implementation may be found in \cite{migflow,henry2025multiscale}.

\section{Numerical methodology}\label{sec:numer}

\begin{figure}[H]
    \centering
    \includegraphics[width=.75\textwidth]{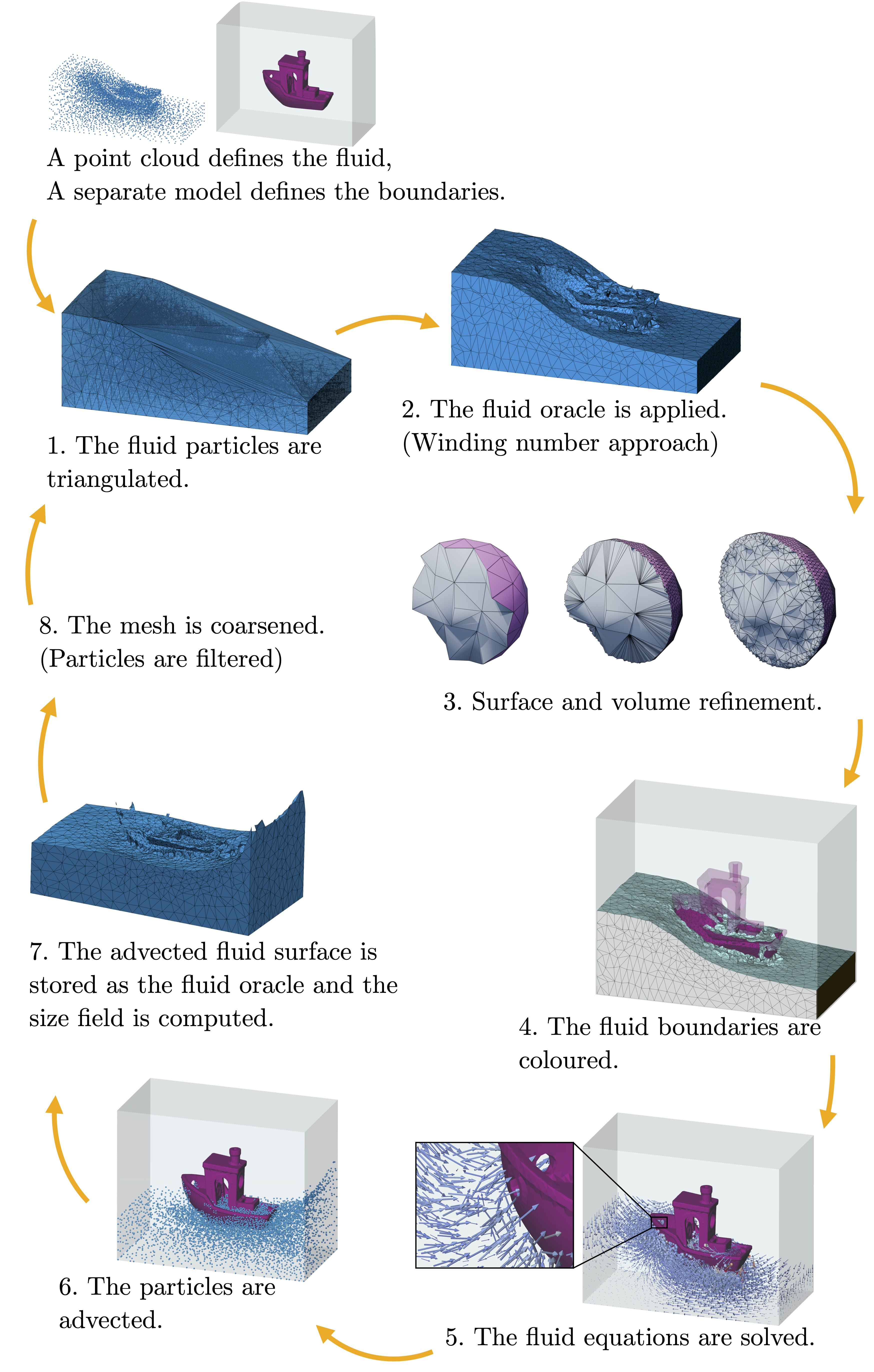}
    \caption{Illustration of the PFEM algorithm.}\label{fig:pfem_scheme_3D}
\end{figure}

The complete algorithm is summarized in Figure~\ref{fig:pfem_scheme_3D}.
First, a point cloud $\mathcal{S}(\mathbf X)$ is generated to represent the initial configuration of the fluid.
A Delaunay tetrahedralization $\mathcal{T}(\mathcal{S})$ is then constructed (Step~1), yielding a mesh of the convex hull of $\mathcal{S}(\mathbf X)$.
To recover the actual fluid domain, it is necessary to identify which elements of $\mathcal{T}(\mathcal{S})$ belong to the fluid (Step~2).
In this work, we propose an approach that improves upon the classical $\alpha$-shape method by incorporating information about the fluid boundaries advected from the previous time step. This aspect is discussed in detail in Section~\ref{sec:oracle}.

Once the domain has been defined, a mesh adaptation step is applied to preserve the overall mesh quality and to enforce the prescribed mesh size field throughout the domain (Step~3). The mesh adaptation procedure is described in Section~\ref{sec:adapt}.

The next step consists of detecting the boundaries in order to apply boundary conditions correctly (Step~4). Section~\ref{sec:bnd} describes in detail the approach adopted in the present work.

The equations of motion are then solved on the mesh (Step~5), and the particles are advected explicitly using Equation~\ref{eq:adv} (Step~6).
The advected fluid surface is subsequently computed (Step~7), which enables the definition of the fluid domain at the next time step.
Finally, a mesh coarsening step is performed to prevent particle clustering (Step~8).
In practice, if two particles are closer than a threshold value determined by the prescribed size field, one of them is removed.

In the following sections, the three contributions proposed in this work are described in more detail.
We first present the improved domain reconstruction method (Section~\ref{sec:oracle}), followed by the adaptive mesh refinement strategy (Section~\ref{sec:adapt}), and finally the boundary management procedure (Section~\ref{sec:bnd}).


\subsection{The fluid oracle}\label{sec:oracle}
A well-known limitation of the particle finite element method, extensively discussed by \cite{franciMass}, is the lack of volume conservation in free-surface flows.
As noted by \cite{falla}, the error in volume conservation has two main sources.

The first source arises from the numerical resolution of the governing equations.
The velocity field is not strictly divergence-free owing to the PSPG stabilization.
In addition, the explicit advection of particles implies that the divergence-free condition is not exactly preserved after the transport step.
Although this issue can be mitigated, for example by employing higher-order time integration schemes such as the second-order method proposed in \cite{LEYSSENS2025114082}, it remains a source of error.

The second source of error is associated with the remeshing procedure, and more specifically with the use of the standard $\alpha$-shape algorithm as an oracle to define the fluid domain and its boundaries.
In its simplest form, the $\alpha$-shape $\mathcal{T}_\alpha(\mathcal{S})$ is defined as the subset of elements of $\mathcal{T}(\mathcal{S})$ whose circumradii are smaller than a prescribed parameter~$\alpha$ \cite{alphashapeSurvey, alphaShapes3D}.
Accordingly, an element $e$ of $\mathcal{T}(\mathcal{S})$ with circumradius $R_e$ belongs to $\mathcal{T}_\alpha(\mathcal{S})$ if $R_e < \alpha$.
In \cite{leyssens}, the authors extended this definition by introducing a spatially varying $\alpha$ parameter.
In this formulation, $\alpha(\mathbf{x})$ is defined as a field, which can be naturally coupled with adaptive mesh refinement.
The criterion remains unchanged, \emph{i.e.}, for each element, the condition $R_e < \alpha(\mathbf{x})$ must be satisfied.

Although, as presented in \cite{leyssens}, the $\alpha$-shape maintains a coherent structure and is relatively stable under small motion of the points, it possesses some drawbacks that introduce unphysical behaviour in the detection of the domain.

\begin{figure}[h!]
    \centering
    \includegraphics[width=.7\textwidth]{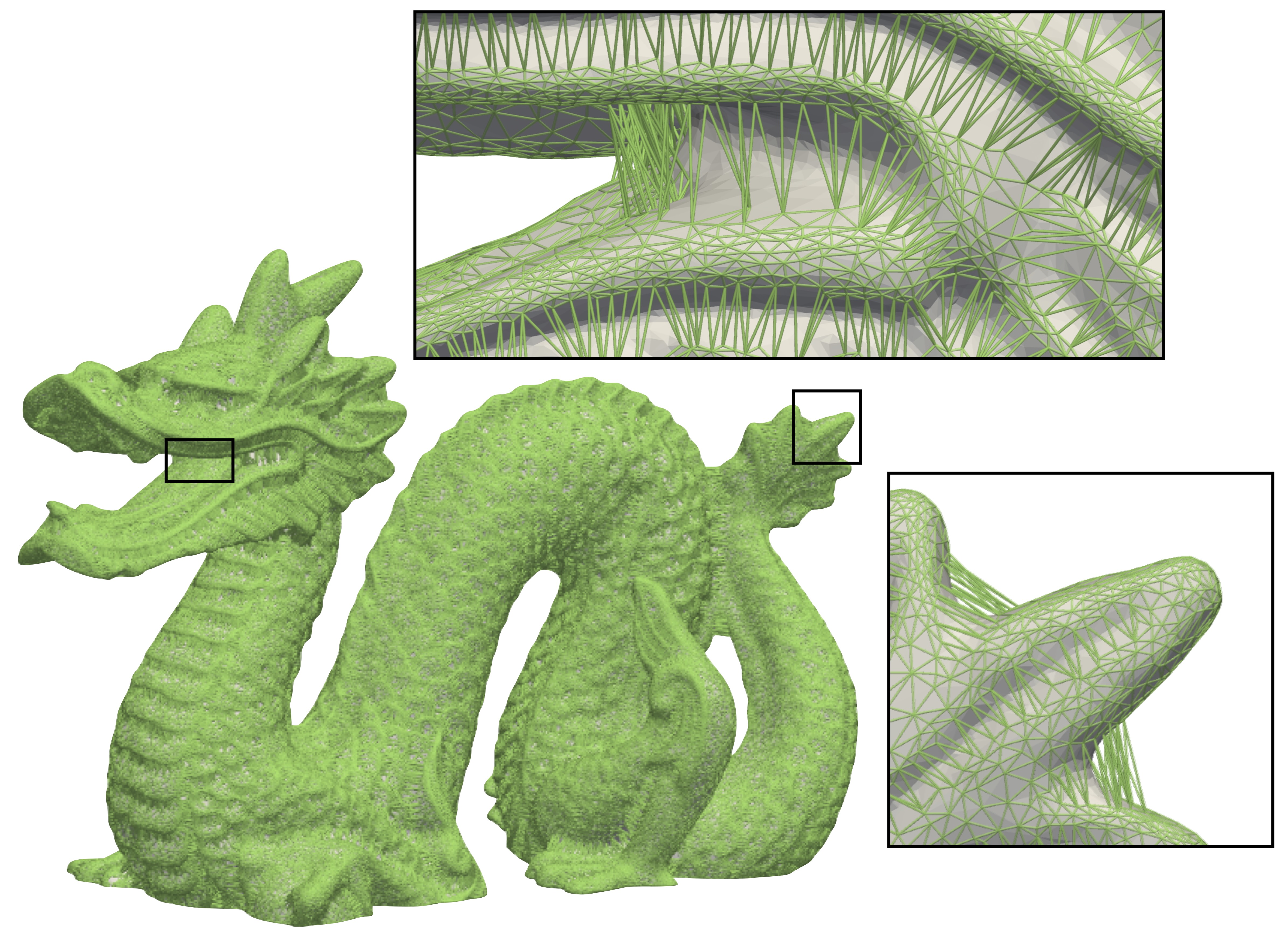}
    \caption{Shape reconstruction of a dragon using the $\alpha$-shape, with the ground truth in grey. The $\alpha$-shape does not capture sharp angles. In this example, $\alpha=1.3$.}\label{fig:dragon}
\end{figure}

These issues have been pointed out in other fields, such as volume reconstruction from point clouds obtained from LiDAR data \cite{imaging_alphashape}.
Two main issues cause these undesirable artefacts. 
First, the $\alpha$-shape does not behave well when points are not homogeneously distributed. 
Holes appear in regions that are less densely concentrated in nodes. 
Second, small angles and sharp features are not well captured. 
Since the algorithm relies only on the shape and size of elements, it has no knowledge of geometrical features. 
Moreover, choosing the value of $\alpha$ is not straightforward, and as pointed out in \cite{franciMass}, this has a significant effect on the results of the simulation.
This is illustrated in Figure \ref{fig:dragon}: small concave angles in the geometry are lost.

\begin{figure}[h!]
    \centering
    \includegraphics[width=.7\textwidth]{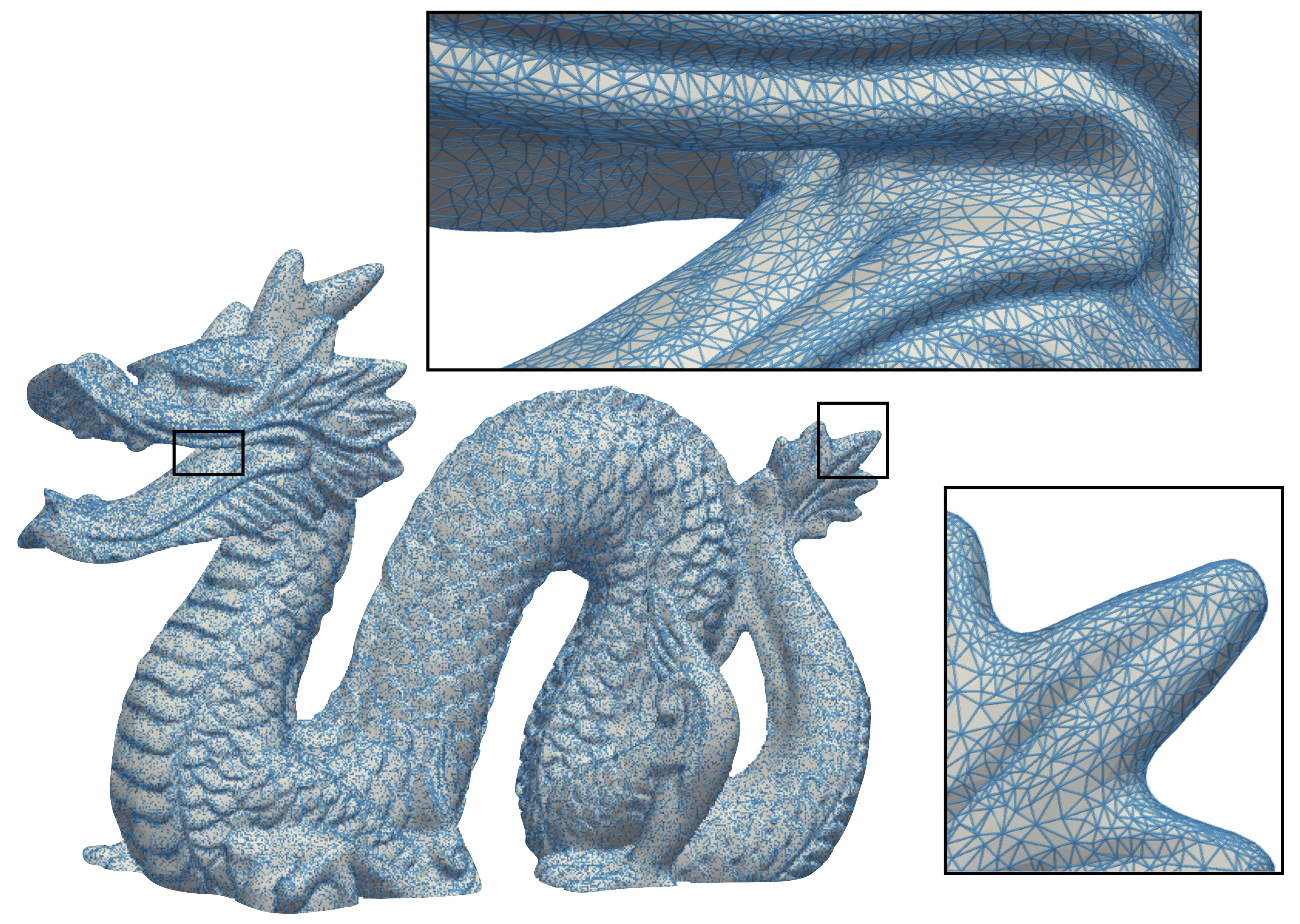}
    \caption{Shape reconstruction of a dragon using the winding number, as proposed by \cite{windingNumber2}, with the ground truth in grey. Sharp angles in the input are well recovered.}\label{fig:dragon_wn}
\end{figure}

We propose to use another kind of fluid oracle to indicate which elements in the convex hull are actually fluid.
We continue to take advantage of the well-defined theoretical properties of the Delaunay triangulation and the algorithmic efficiency of the PFEM framework, while enriching it with the physical information available in the simulation.

To this end, we seek a method capable of determining whether a given element lies inside or outside the fluid domain.
Once this classification is available, elements that are in fact outside the domain can be removed.
A particularly well-suited approach for this purpose is the winding number algorithm introduced in \cite{windingNumber1}.
In its basic formulation, the winding number computes the number of times a closed, oriented boundary wraps, or \emph{winds}, around a point~$p$.



%

\begin{figure}[h!]
    \centering
    \includegraphics[width=.4\textwidth]{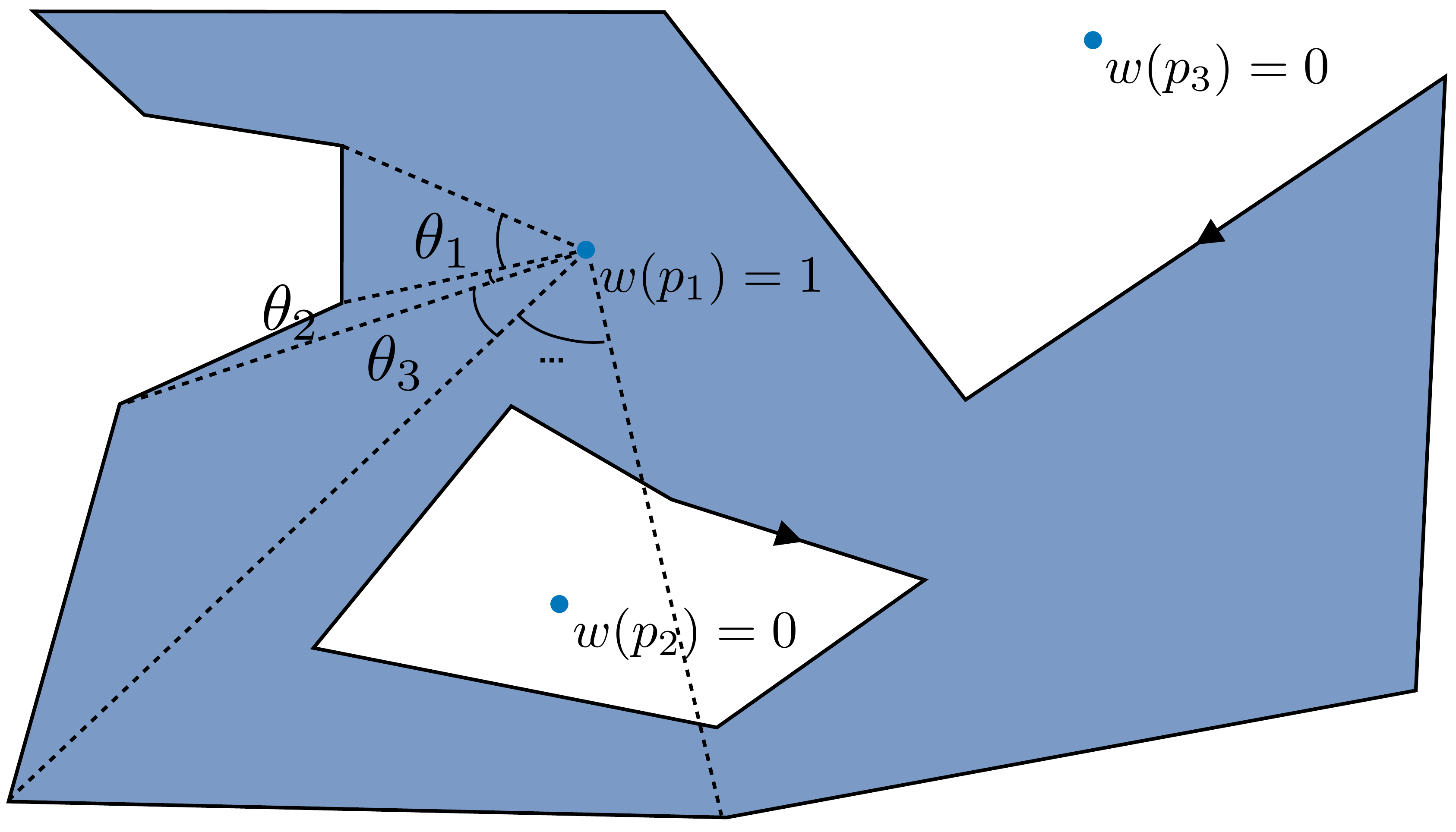}
    \caption{The winding number $w(p)$ is equal to 1 inside an oriented closed curve, and 0 outside.}\label{fig:winding}
\end{figure}

Though it is general for any closed Lipschitz curve, for piecewise linear curves, such as the one in Figure \ref{fig:winding}, it can be computed as the sum of angles around the query point.
For a closed curve in 2D, the sum of angles formed by the two points of a bounding edge and point $p$ is equal to $2\pi$ if $p$ is inside, and $0$ if $p$ is outside.
The only requirement is that the edges are oriented.
If there is a hole in the polygon, the result does not change. 
For instance, point $p_2$ in Figure \ref{fig:winding} is outside: the sum of angles of the inner loop (the hole) is $-2\pi$, and the outer loop is $2\pi$. The sum is $0$. 

\begin{align}
    w(p) = \frac{1}{2\pi}\sum_i \theta_i
\end{align}

The result is identical in 3D, by considering the sum of solid angles formed by the triangles of a closed surface. 
We use an efficient implementation of the winding number, proposed in \cite{windingNumber2}.
Indeed, computing this winding number may seem an expensive process for large meshes, especially if the number of query points is high.
However, the implementation of \cite{windingNumber2} is very efficient thanks to the use of multipole decompositions to simplify evaluations with boundary edges that are far from the query point. 
Figure \ref{fig:dragon_wn} shows that the identification of the boundaries of an object and topological changes are much more precise with the winding number than with the $\alpha$-shape.
Moreover, the implementation of \cite{windingNumber2}, named the generalized winding nubmer, has been implemented for point clouds and soups of triangles. 
It therefore still provides a robust in/out answer even for defects such as self-intersections. 

\begin{figure}[H]
    \centering
    \includegraphics[width=.65\textwidth]{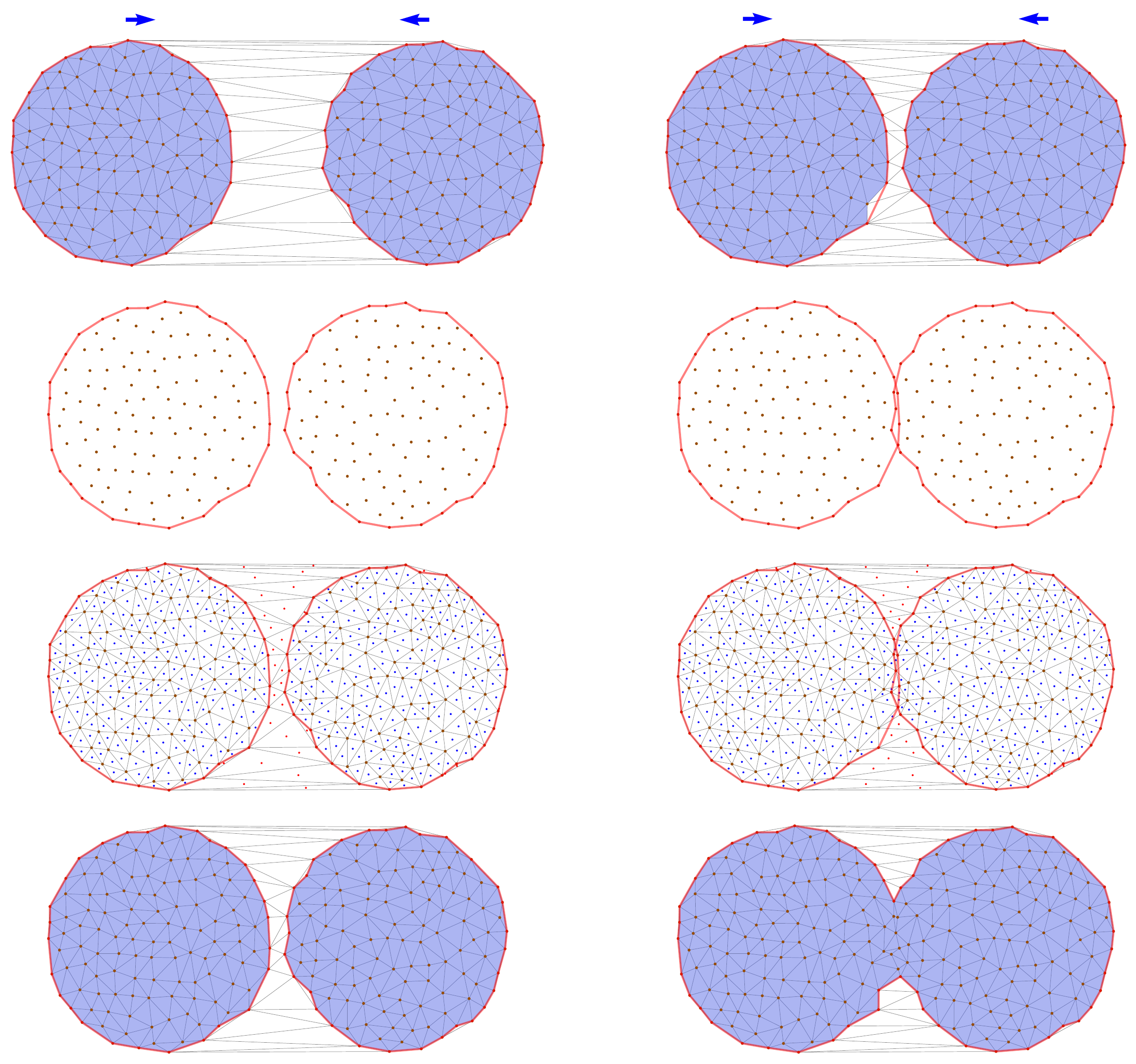}
    \caption{Domain definition with winding number approach. The figures represent two time steps (left and right). From the previous time step, the boundary is advected using the velocity field. We then re-triangulate the nodes, and detect, for each triangle, whether its barycenter lies inside the closed volume defined by the advected boundary. We then recover the shape.}\label{fig:winding_number_process}
\end{figure}

The approach to detect the new domain within a PFEM simulation is as follows.
The winding number requires an input, closed surface for which it can define the inside from the outside. 
In the PFEM algorithm, the fluid oracle is called upon when we have re-generated the Delaunay triangulation.
Referring to Figure \ref{fig:pfem_scheme_3D}, this corresponds to step (3). 
With the classical $\alpha$-shape procedure, the topological information of the mesh, and therefore also the boundaries, of the previous time step are lost.

Instead, to be able to properly use the winding number, the boundaries of the fluid are advected using the velocity of the previous time step, and this is considered as the closed surface on which the winding number may be computed at the new time step.
The barycenter of each tetrahedron in the Delaunay mesh is computed, and this point is used to compute the winding number with respect to the advected fluid surface of the previous time step. 
The procedure for two time steps is summarized in Figure \ref{fig:winding_number_process}.
This introduces a physical notion into the computation of the fluid domain, which cannot be achieved with the classical $\alpha$-shape approach. 

\subsection{Adaptive refinement}\label{sec:adapt}
Different methods for mesh adaptation in PFEM have been proposed in 2D. 
In \cite{falla}, an extensive study has been performed to understand the phenomena causing elements to be wrongly removed from the simulation, and an approach based on edge-splitting was proposed to perform adaptive refinement. 
These results showed improvement in terms of mass conservation, and accurate tracking of topological changes. 
More recently, a Delaunay-based mesh refinement approach was proposed by the authors in 2D \cite{leyssens}.
The method focuses on maintaining quality elements using a strategic node insertion at element circumcentres. 

In 3D, mesh adaptation has not yet been addressed much in the context of PFEM. 
One approach proposed in the literature is the concept of mesh smoothing \cite{meduri2019}.
This is especially relevant in explicit time integration schemes, as the main objective is to remove sliver elements to maintain a reasonable time step. 


It is also worth noting the work of \cite{rizzieri}, who proposed a de-refinement procedure, specific to their studied case of 3D concrete printing. 
The process consists of removing nodes in regions where the dynamics are of less importance.
Using a geometrically defined criterion (the distance to the printing nozzle), a size field allows particles to be removed throughout the simulation to maintain reasonable computational costs.
To address the quality of elements, and more specifically the issue of slivers in three-dimensional simulations using explicit time integration, PFEM has recently been coupled with the virtual element method \cite{PVEM}, by creating an arbitrary polyhedral element around slivers and seamlessly connecting them to the finite element tetrahedra of the mesh.

In implicit time integration schemes, the presence of isolated slivers is not an issue.
As demonstrated in \cite{kuvcera2016necessary} and \cite{hannukainen2012maximum}, the maximum angle condition is not a necessary condition for convergence of finite element problems. 
The problem of multiple slivers concentrated together to form bands, however, may lead to locking phenomena and therefore impact the solution. 
Yet, recent work by \cite{quiriny2024tempered} showed that it is still possible to perform simulations, namely using the Tempered Finite Element Method, to avoid locking the solution.

For different reasons, it is often desirable to have local refinement in a simulation domain. 
Indeed, a uniform mesh size is often not optimal, both in terms of computational cost and accuracy. 
One may be interested in the flow in the vicinity of an obstacle, near a boundary, or at the free surface. 
In other cases, the physics often requires more refinement in regions with high gradients. 
It turns out that the PFEM is very well suited for local refinement and non-uniform meshes, since mesh adaptation is at the core of the method. 

Mesh adaptation cannot have any influence on the shape of the domain. 
In other words, once the fluid domain has been defined, mesh refinement should not alter its shape. 
Yet in the case of free surface flows, for instance, most topological changes occur at the boundaries, so we want to be able to allow refinement on the free surface. 
The work presented in \cite{leyssens}, which considered mesh refinement for two-dimensional applications, used Chew's algorithm for mesh refinement \cite{chew}. 
In 3D, however, mesh refinement becomes more complex, since the boundary itself becomes a surface.


The approach proposed in this work is to divide the refinement process in two steps. 
First, we perform a mesh refinement of the surface elements through an edge splitting procedure. 
Through this process, the surface is refined and respects the size field, but its shape is not altered.
Then, the volume elements are refined through node insertion without changing the surface. 

\subsubsection*{Surface refinement}

\begin{figure}[h!]
    \centering
    \includegraphics[width=\textwidth]{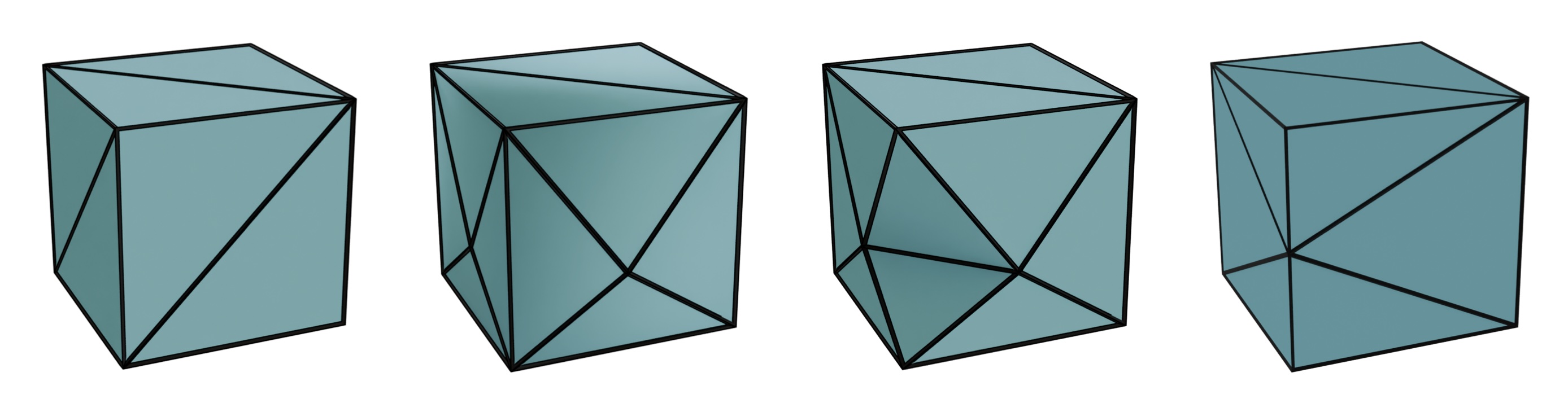}
    \caption{From left to right: initial mesh of the cube; node insertions inside an element; flipped edge; node insertion on an edge.}\label{fig:cube_flipped}
\end{figure}

Once the surface of the fluid has been defined, any changes made to the mesh should not alter its shape. 
An algorithm that includes edge flips can therefore not be used, as illustrated in Figure \ref{fig:cube_flipped}.
The first three images show that, if points are inserted inside triangular elements, and edge flips are then performed, we alter the shape of the domain.

\begin{figure}[h!]
    \centering
    \includegraphics[width=.6\textwidth]{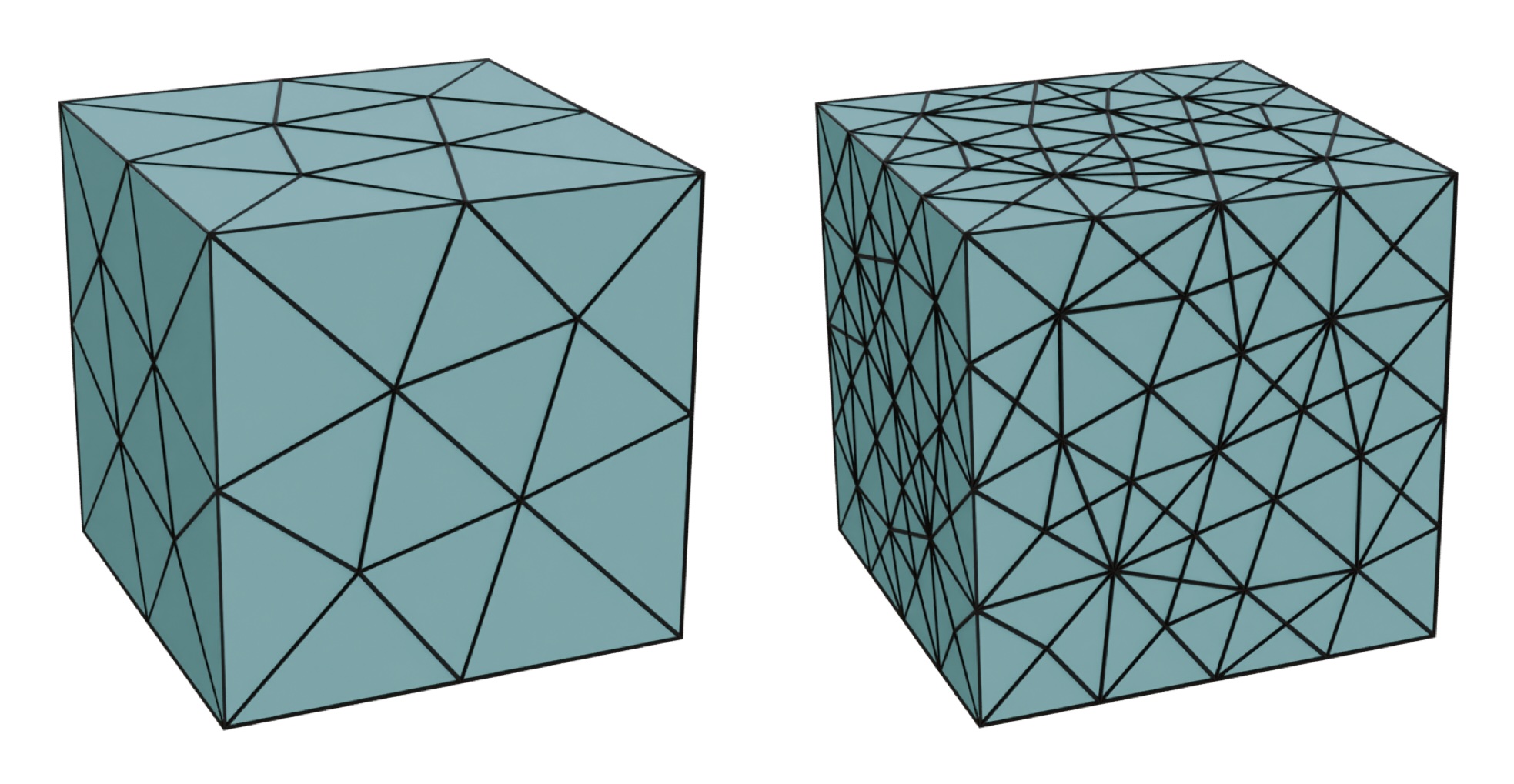}
    \caption{The edge splitting approach guarantees that the geometry of the surface is maintained, while improving mesh quality.}\label{fig:cube_edge_split}
\end{figure}

Another approach, which guarantees that the shape is maintained while still improving quality, is the longest edge-splitting method. 
As can be seen in the fourth image of Figure \ref{fig:cube_flipped} and Figure \ref{fig:cube_edge_split}, the cube's shape is maintained.

Longest-edge splitting is a well-known mesh refinement and improvement technique in the meshing community. 
For instance, \cite{rivara84, rivara97} presented algorithms for edge splitting. 
The main advantages, besides maintaining the shape, are the relative ease of implementation and the tendency to improve the quality. 
Although there is no strong mathematical proof that edge splitting improves mesh quality, it has been shown, for instance in \cite{rosenberg_proof_bisect} that the elements produced by the edge splitting approach have angles that are bounded by the angles of the input triangulation. 
In other words, no small angles are generated by the longest-edge splitting algorithm. 

Although different versions of the longest edge splitting method have been proposed, the one used in this work is largely inspired by the second algorithm presented in \cite{rivara84}.
The principle is very straightforward: while there are long edges in the mesh, find the longest one and split it at its midpoint to generate two new elements. 

Let us define the surface triangulation, obtained after reconstructing the shape of the domain, as $\partial \mathcal{T}(\mathcal S)$, and an edge $\tau_{ij}$ as the $j$-th edge of element $i$ from $\partial \mathcal{T}(\mathcal S)$ between nodes $\text v_m$ and $\text v_n$. 
$l(\tau_{ij})$ is the length of $\tau_{ij}$, and $h(\text v_m)$ is the value of the mesh size field at $\text v_m$.
The algorithm can be summarised as follows: 

\begin{algorithmic}
    \While {$\partial \mathcal{T}(\mathcal S)$ contains long edges}
        \State Sort $\tau_{ij}$ from longest to shortest
        \State Get $\tau^1_{ij}$, the longest edge in $\tau_{ij}$
        \If {Split\_allowed($l(\tau^1_{ij})$) $\And$ $l(\tau^1_{ij}) > 0.5(h(\text v_m)+h(\text v_n))}$
            \State Split $\tau^1_{ij}$
            \State Add the new edges to $\partial \mathcal{T}(\mathcal S)$ and in the sorted list
        \Else
            \State Remove $\tau^1_{ij}$ from the sorted list
        \EndIf
    \EndWhile
\end{algorithmic}

\begin{figure}[H]
    \centering
    \includegraphics[width=.4\textwidth]{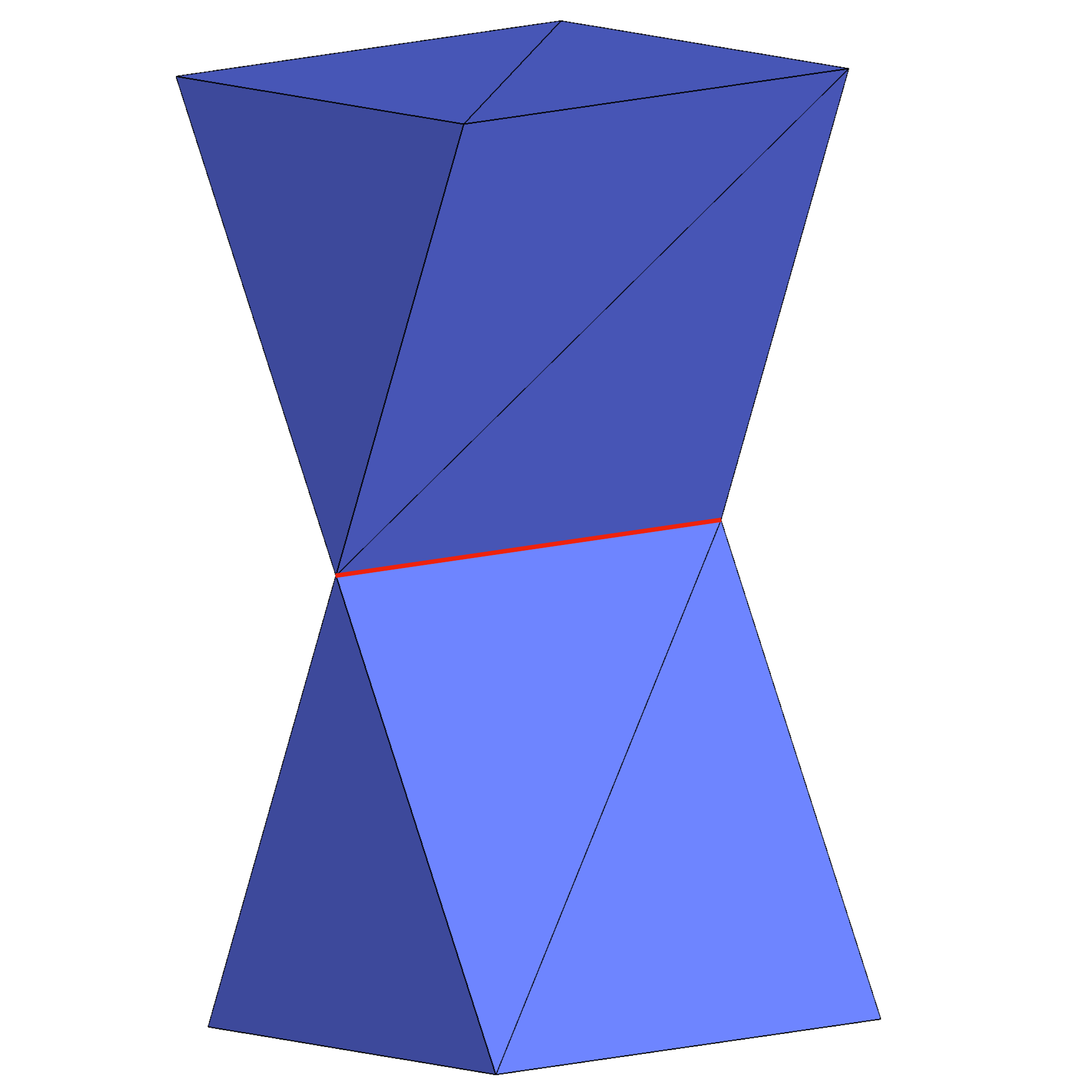}
    \caption{The edge highlighted in red is a non-manifold edge: it is connected to four triangles of the boundary. This means that, in a mesh data-structure such as a half-edge, it is actually present twice. Splitting this edge is therefore numerically error-prone.}\label{fig:non-manif}
\end{figure}

We employ a half-edge data structure \cite{halfEdge} to perform these operations on the surface mesh.
The reason we need to check that the edge can actually be split is that, in the case of a topologically complex mesh such as the one resulting from free surface simulations, it often happens that the surface is non-manifold, \textit{i.e.}, it contains edges that are connected to more than two triangles, as illustrated in Figure \ref{fig:non-manif}.
In this case, splitting a non-manifold edge requires considerably more work and specific data structures to ensure robustness.
In the current implementation, we have made the choice to avoid splitting these edges. 

\subsubsection*{Volume refinement}

The surface boundary is now clearly defined and respects the size field. 
This can be given as input to refine the internal volume mesh. 
In other words, a Delaunay refinement process can be performed on an input comprised of all the particles of the fluid and the triangular surface mesh. 
The process then starts by generating the constrained Delaunay tetrahedralization of the nodes, with the surface triangles as constraints. 

\begin{figure}[H]
    \centering
    \includegraphics[width=\textwidth]{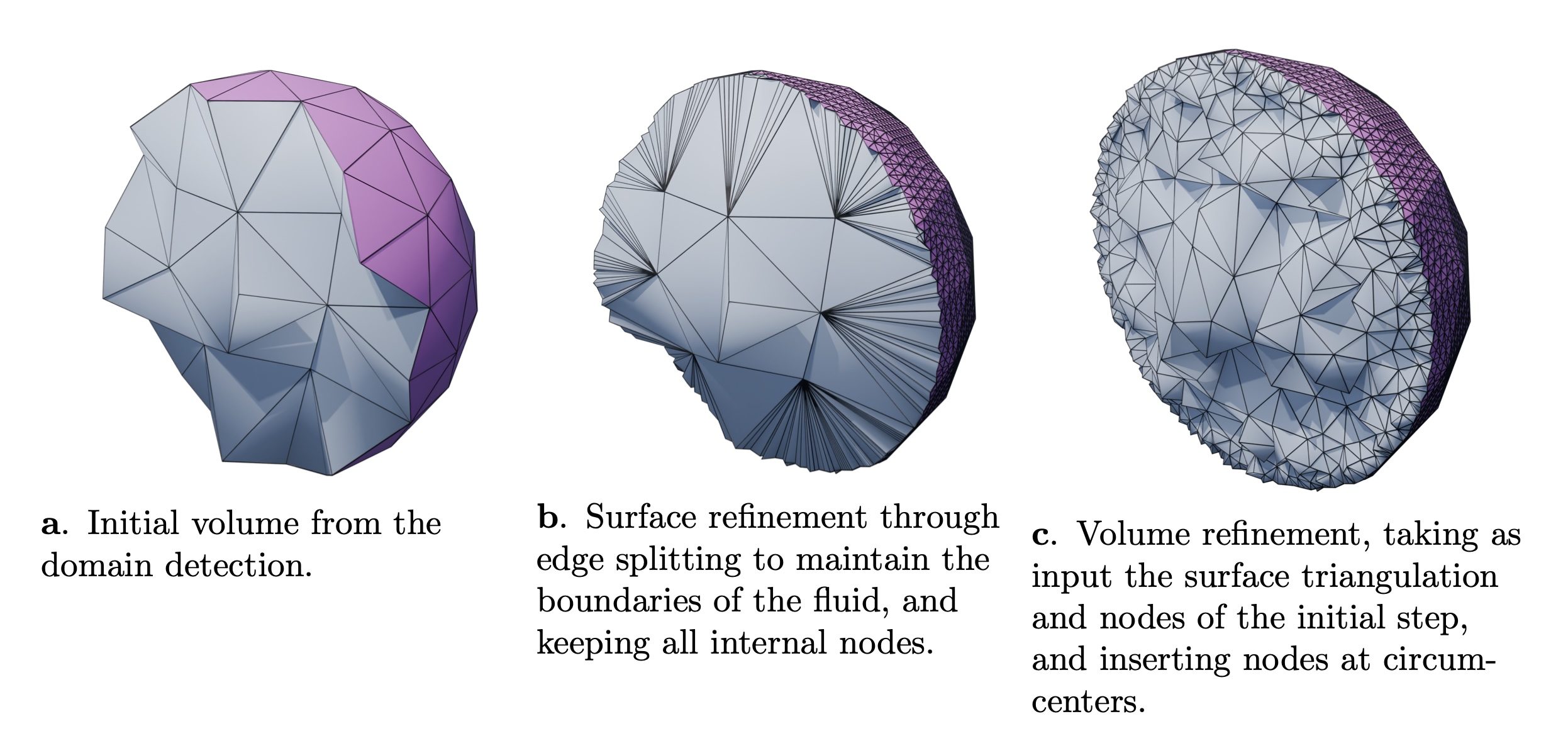}
    \caption{Illustration of the refinement procedure, composed of two steps: surface refinement, followed by volume refinement.}\label{fig:refine}
\end{figure}

We use the parallel algorithm proposed in \cite{celestin} to perform the volume mesh refinement. 
Figure \ref{fig:refine} illustrates the three steps.

The element refinement follows a generalization of the algorithm of \cite{chew} for triangular mesh refinement in the plane, to three-dimensional tetrahedral meshes. 
This process has also been implemented for instance in Tetgen \cite{tetgen}.
The tetrahedra of the mesh are progressively refined in the following manner.
First, vertices are created at the circumcenter of elements whose circumradius is significantly larger than the desired mesh size. 
Next, a filtering procedure verifies whether the new points are not too close to each other or to existing nodes.
This is necessary to avoid the creation of small edges. 
Finally, points are inserted into the mesh through a Delaunay algorithm. 
Each of these three processes is implemented in parallel. 
For further details, the interested reader is referred to \cite{celestin} and \cite{marotThese}.
Marot's implementation of this process is available in Gmsh \cite{gmsh}.

\subsubsection*{Mesh coarsening}

Mesh coarsening is an important step within an adaptive refinement context, as we do not want to have an ever-increasing number of particles within the simulation. 
For efficiency reasons, this de-refinement step is performed before triangulating the nodes again (step 8 in Figure \ref{fig:pfem_scheme_3D}).
It is best to perform this de-refinement step before triangulating the nodes again, as this avoids having to generate a new mesh twice.


The particle filtering process is quite straightforward. 
If two particles are too close to each other with respect to the prescribed size field in this region, one of them is filtered out. 
In this naive approach, no special distinction is made between boundary particles or particles from the bulk. 
This deserves more attention, for instance by ensuring that no boundary particles responsible for geometric features in the domain are removed.
The use of an intelligent decimation algorithm, for instance \cite{schroeder1992decimation}, would be an interesting approach.

\subsection{Boundary management}\label{sec:bnd}
In the proposed PFEM approach, the solid geometry is fully decoupled from the fluid representation. 
Hence, to apply the different boundary conditions, each boundary element of the fluid must be assigned the correct boundary: a solid wall or a free surface. 
This must be redefined at each time step due to the geometric and topological evolution of the domain. 

\begin{figure}[H]
    \centering
    \includegraphics[width=\textwidth]{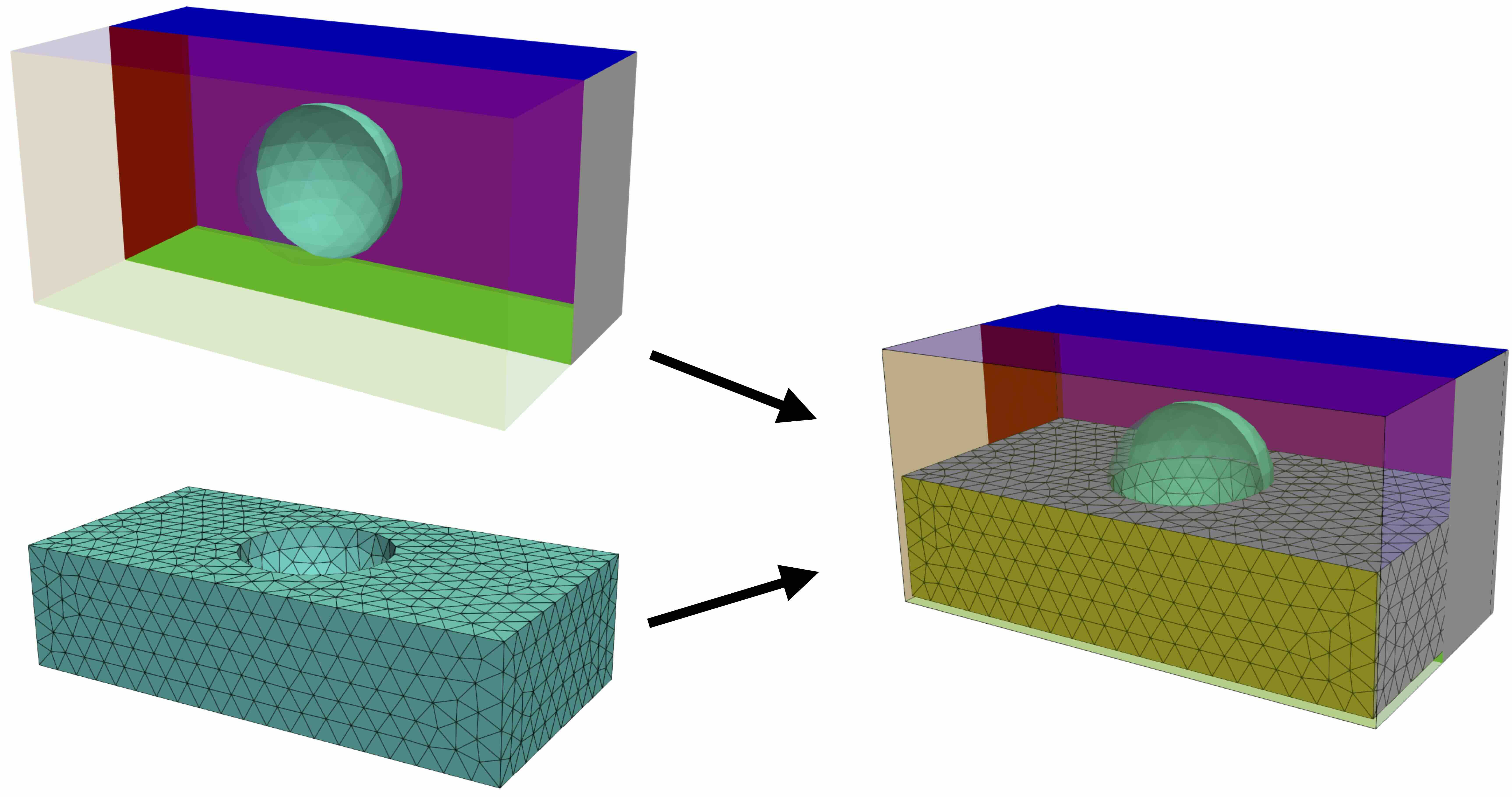}
    \caption{Illustration of the boundary colouring procedure: the boundary faces of the fluid (left, below) are coloured according to the geometry of the wall (left, top).}\label{fig:colouring}
\end{figure}

A boundary \textit{colouring}, or identification, is therefore a necessary step in the algorithm (step 4, Figure \ref{fig:pfem_scheme_3D}). 
Figure \ref{fig:colouring} illustrates the colouring procedure. 

To achieve this colouring step, we define an abstract boundary manager. 
This abstract boundary manager must be able to perform the following processes: 
\begin{enumerate}[label=(\alph*)]
    \item \textbf{In/Out detection.} Detect whether a point lies inside or outside the physical domain.
    This is required during the element filtering step to determine the shape of the fluid (step 3 of Figure \ref{fig:pfem_scheme_3D}). 
    Some elements may be generated outside the geometrical domain. 
    This is more often the case for non-convex boundaries. 
    In this case, the elements whose centers of mass are outside the domain are removed.
    \item \textbf{Segment-boundary intersection.} Intersect a segment with the boundary. 
    This is needed to project fluid particles back to the wall if they cross a boundary during the advection step. 
    In this case, their displacement is adjusted such that they are blocked by the wall. 
    \item \textbf{Point-to-boundary projection.} Find the closest point on the boundary from another point.
    This is required to colour the boundary triangles. 
    If the distance from the center of mass of a boundary triangle to a wall is below a given threshold, typically 1\% of the mesh size, the triangle is coloured with the colour of the wall. 
    Otherwise, it is considered a free surface.
\end{enumerate}

As long as these three processes can be defined, any approach can be implemented to colour boundaries and subsequently apply proper boundary conditions.
For simple cases, it is possible to define these processes analytically. 
For complex shapes, however, the analytical approach is no longer possible. 
We therefore define a boundary triangulation, which represents the boundary with linear elements.
The elements of the boundary are then stored in a search data structure, an octree \cite{octree}.
Intersections and projections can then be performed between the points, segments, and triangles. 
The in/out detection is performed with the same algorithm as for the fluid oracle, namely a winding number approach \cite{windingNumber2}.

\subsubsection*{Boundary particles re-projection}

\begin{figure}[H]
    \begin{center}
        \includegraphics[width=.6\textwidth]{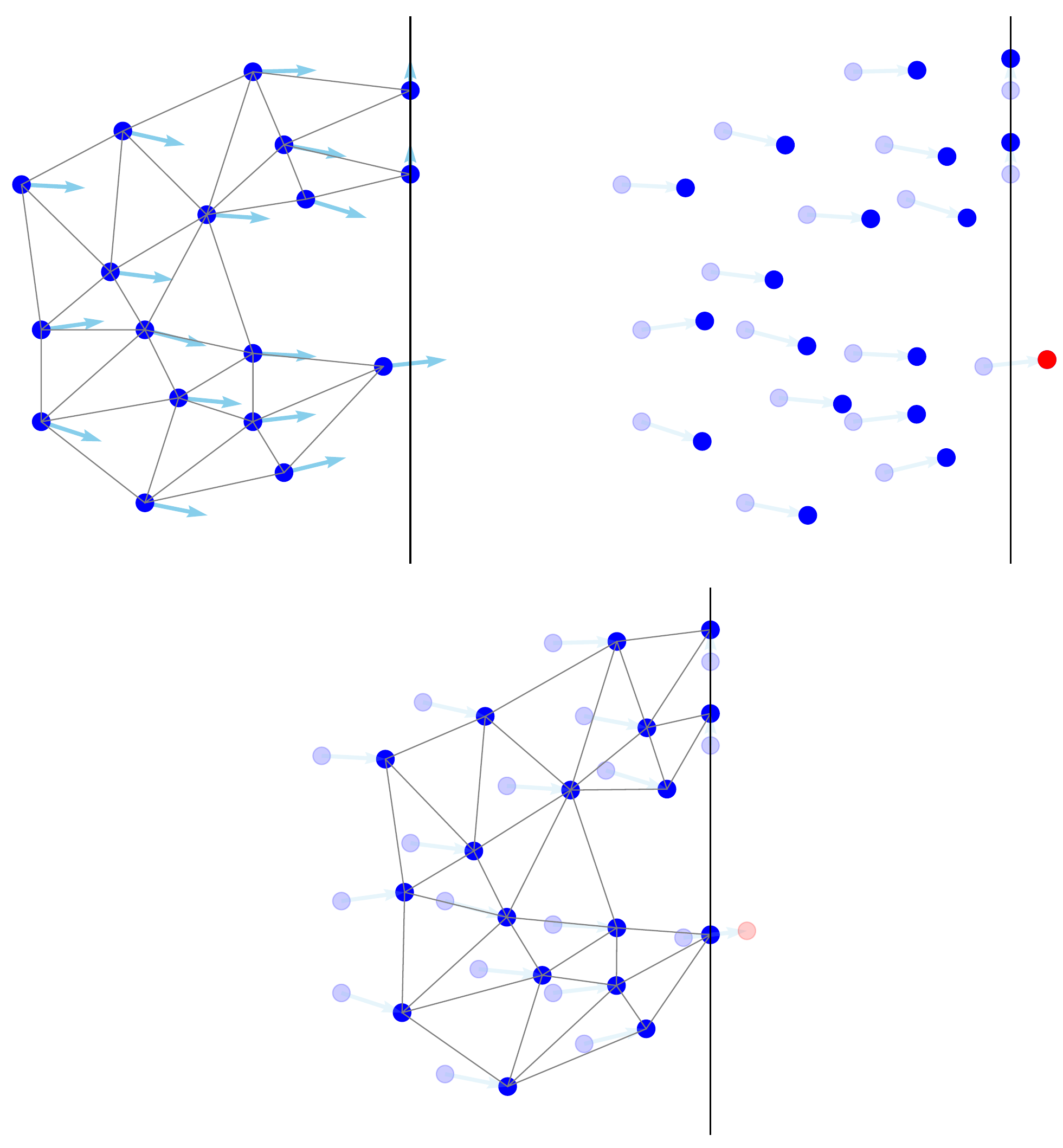}
        \caption{The position of the particles that cross a wall during the advection step, for example the particle in red in this figure, are corrected by projecting them on the wall.}\label{fig:reproj}
    \end{center}
\end{figure}

In PFEM, the displacement of the fluid particles is performed by explicitly integrating the advection equation \ref{eq:adv}.
For each particle $i$, the position is updated as follows:
\begin{align}
    \mathbf x_i^{t+1} = \mathbf x_i^t + \mathbf u_i^{t+1}\Delta t.
\end{align}
Due to this explicit displacement, particles may cross boundary walls as they do not know in advance about the presence of walls. 
A correction is made to these particles' velocities to project them onto the boundary, as illustrated in Figure \ref{fig:reproj}.

Another correction concerns particles whose face has been detected as part of a boundary.
For instance, a triangle may have been coloured as a wall because it is close enough to that wall, within the tolerance of 1\%. 
In that case, we project the particles of that triangle onto the solid boundary geometry, so that the fluid domain conforms to the geometry. 


\section{Validations}\label{sec:valid}
We present different test cases to illustrate the proposed method. 
The first is an analytical case, while the following two are performed in simple geometries to show consistency and to compare with experimental and numerical results from the literature.

\subsection{Vortex-in-a-box}

\begin{figure}[H]
    \centering
    \includegraphics[width=\textwidth]{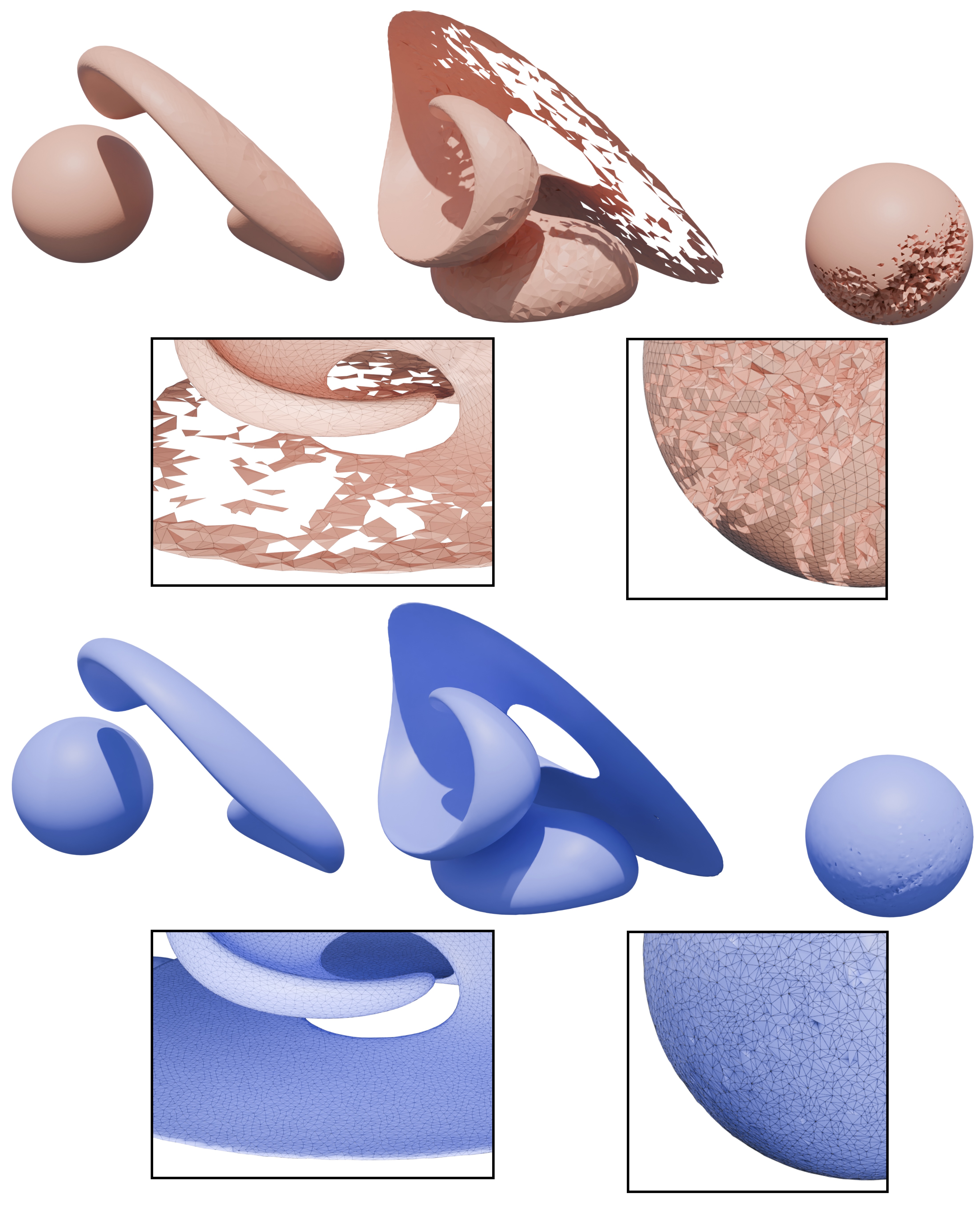}
    \caption{Vortex-in-a-box without (top) and with (bottom) refinement.
    Times: 0s, 1.0s, 2.0s, 4.0s}\label{fig:vortex-in-a-box}
\end{figure}

To illustrate the refinement procedure as well as the domain detection algorithm based on the winding number independently from the fluid solver, we consider an analytical velocity field, initially proposed by \cite{vortex3D}.
This well-known test case checks the conservation properties of a method in convection-dominated problems by applying an incompressible, rotating velocity field that leads to increasingly small features.
The considered domain is a unit cube.
Initially a sphere of radius 0.15 is positioned at $(0.35, 0.35, 0.35)$. 
The given velocity field is the following: 
\begin{linenomath}
    \begin{align}
        u_x &= 2 \text{sin}^2 (\pi x) \text{sin}(2\pi y)\text{sin}(2 \pi z)\text{cos}(\pi t/T)\\
        u_y &= -\text{sin}(2 \pi x)\text{sin}^2(\pi y)\text{sin}(2 \pi z)\text{cos}(\pi t/T)\\
        u_z &= -\text{sin}(2 \pi x)\text{sin}(2 \pi y)\text{sin}^2(\pi z)\text{cos}(\pi t/T).
    \end{align}
\end{linenomath}
The time-dependent factor leads to an inversion of the velocity field after half a period $T$.
Hence, after a full period, an accurate method should return as closely as possible to the initial sphere. 
Figure \ref{fig:vortex-in-a-box} presents a few snapshots for a period $T=4$ $[\mathrm{s}]$, with and without the refinement procedure. 
Table \ref{tab:vortex} shows that, for a similar average number of nodes, the improvement in terms of volume conservation is very significant.

Without refinement, no node insertion is performed, but the nodes are tetrahedralized again at each time step. 
If a node is no longer connected to any element, it is removed from the simulation.
This explains the loss of volume over the course of the simulation. 

With refinement, on the contrary, the mesh adaptation counteracts the stretching of the elements by inserting nodes when elements become too distorted. 
Both surface edge splitting and volume refinement are applied.
Mesh coarsening is also performed by filtering out nodes that are too close to each other.
The coarsening step is important in this test case when $T < t < 2T$, because, as the domain returns to a sphere, the surface-to-volume ratio decreases again.


\begin{table}[h!]
\centering
    \begin{tabular}{| c | c | c |}
        \hline
                       & No refinement & Refinement \\
        \hline
         Number of nodes      & 27372          & 26335       \\
         Volume change & -11.6 \%      & -0.14 \%    \\
         \hline
    \end{tabular}
\caption{Vortex-in-a-box simulation results with and without the refinement algorithm.}\label{tab:vortex}
\end{table}

\subsection{Viscous drop}

In this second test case, we reproduce the three-dimensional version of a known PFEM test case that has previously been studied in 2D by, for instance, \cite{franciMass} and \cite{falla}.
Naturally, the three-dimensional version cannot be compared directly with the two-dimensional version in terms of physics, as the 2D geometry should in fact be assimilated to an infinitely long cylinder instead of a spherical drop. 
Nevertheless, it is still a relevant simulation to study the effect of refinement on the overall result. 
In particular, we investigate the improvement in terms of mass conservation as the mesh is refined. 

\begin{figure}[H]
    \centering
    \includegraphics[width=.7\textwidth]{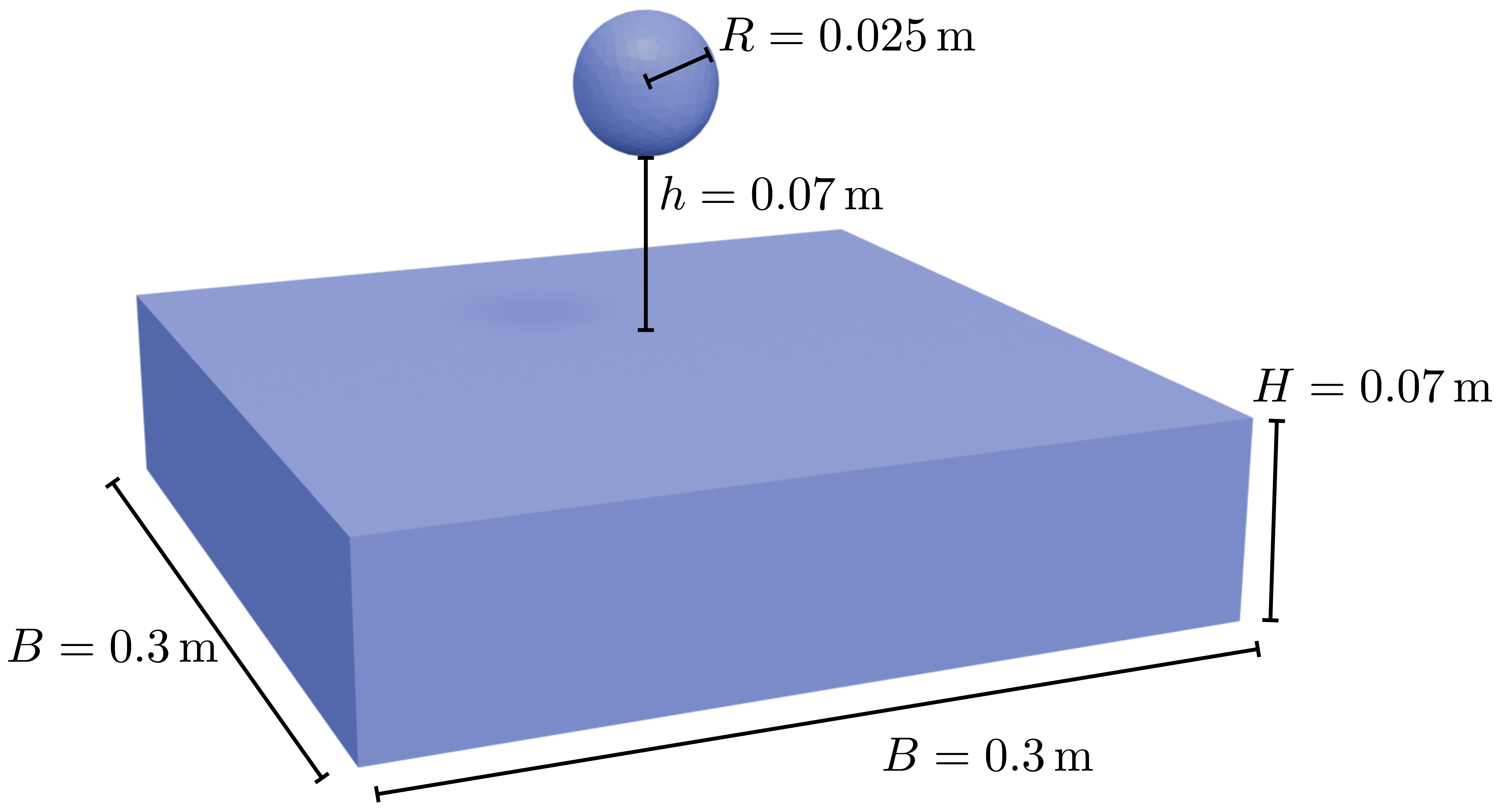}
    \caption{Initial setup of the viscous drop test case.}\label{fig:drop_setup}
\end{figure}

The initial setup of the simulation is presented in Figure \ref{fig:drop_setup}.
The fluid has a viscosity of 0.1 $[\mathrm{Pa\,s}]$, and a density of $10^3$ $[\mathrm{kg\,m^{-3}}]$.

\begin{figure}[H]
    \centering
    \includegraphics[width=\textwidth]{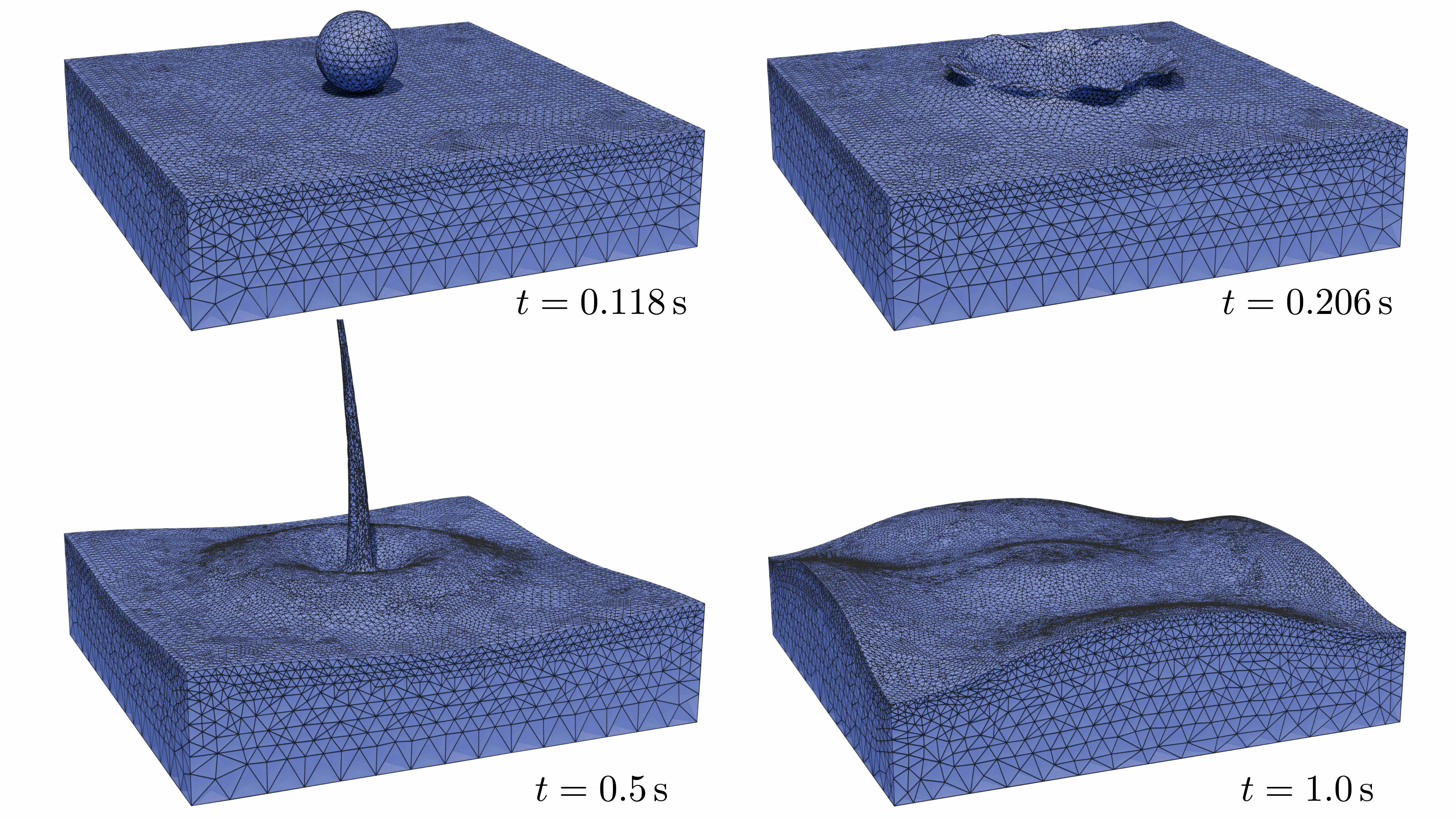}
    \caption{Snapshots of the falling drop experiment.}\label{fig:drop_snapshots}
\end{figure}

Figure \ref{fig:drop_snapshots} presents a few snapshots of the simulation of the falling drop experiment. 
In particular, it is interesting to see that, at $t=0.118$ $[\mathrm{s}]$, as the drop reaches the bulk, the topological change is accurately captured thanks to the local refinement along the free surface, and the winding number approach to track the shape of the fluid domain. 
At $t=0.206$ $[\mathrm{s}]$, the thin film of fluid resulting from the splash is also maintained. 
The loss of radial symmetry can be explained by the non-symmetrical mesh, which could be reduced by further refining the mesh. 
At $t=0.5$ $[\mathrm{s}]$, a high jet-like structure is formed by the high level of velocity reached by the fluid as it fills up the hole generated by the drop. 
Finally, at $t=1.0$ $[\mathrm{s}]$, the fluid reaches an oscillatory state, which is slowly decaying but maintained by the presence of the free-slip boundary condition along the solid walls.

\begin{figure}[H]
    \centering
    \includegraphics[width=\textwidth]{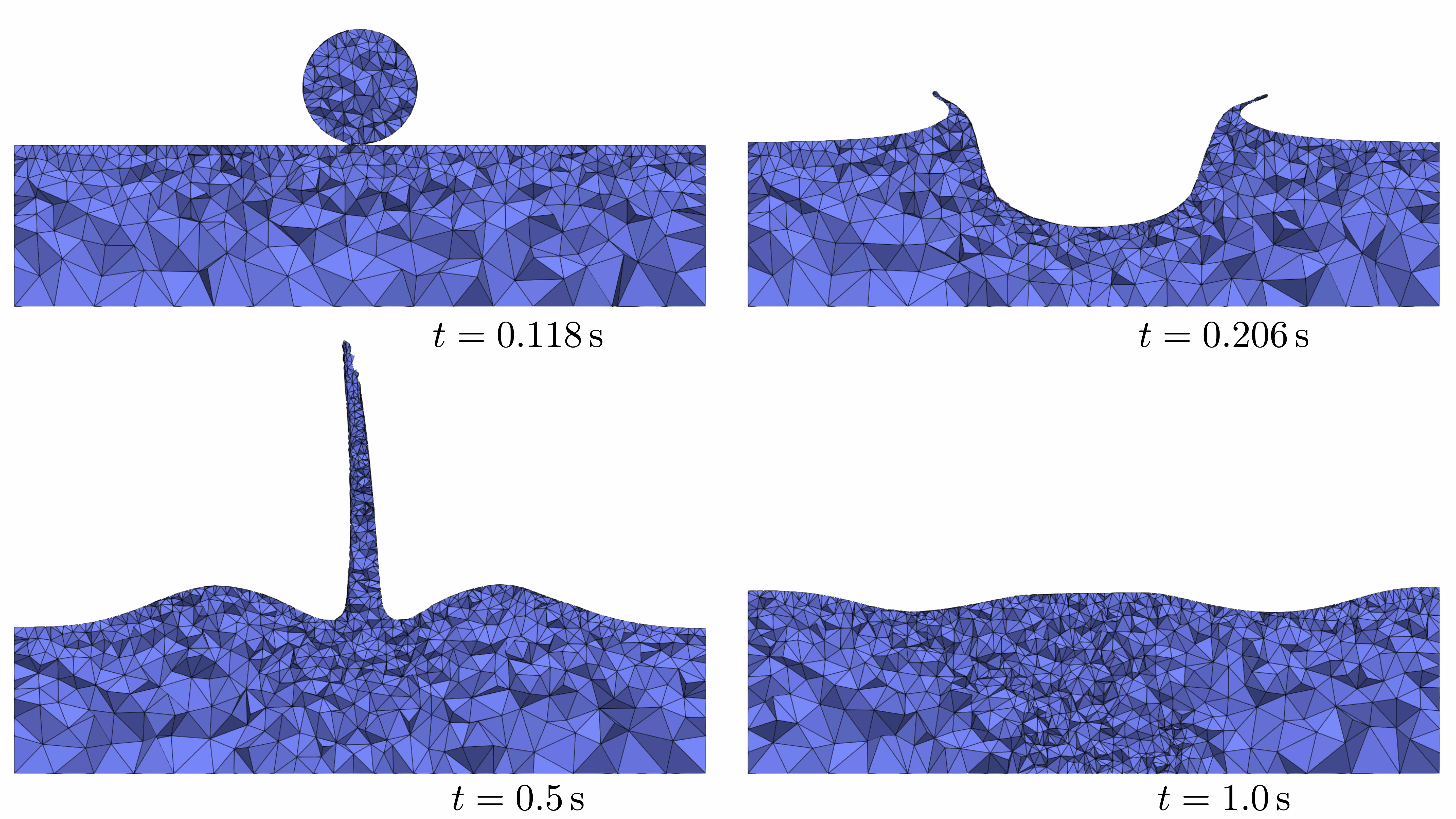}
    \caption{Slices of the falling drop experiment.}\label{fig:drop3D_slices}
\end{figure}

Figure \ref{fig:drop3D_slices} presents a few sliced snapshots of this same three-dimensional simulation. 
The main strength of our approach is the capacity to improve the detection of elements that are inside or outside the fluid. 
The classical $\alpha$-shape approach does this only by evaluating element size and quality, and we add to this the physical information of the advected domain.

\begin{figure}[H]
    \centering
    \includegraphics[width=.8\textwidth]{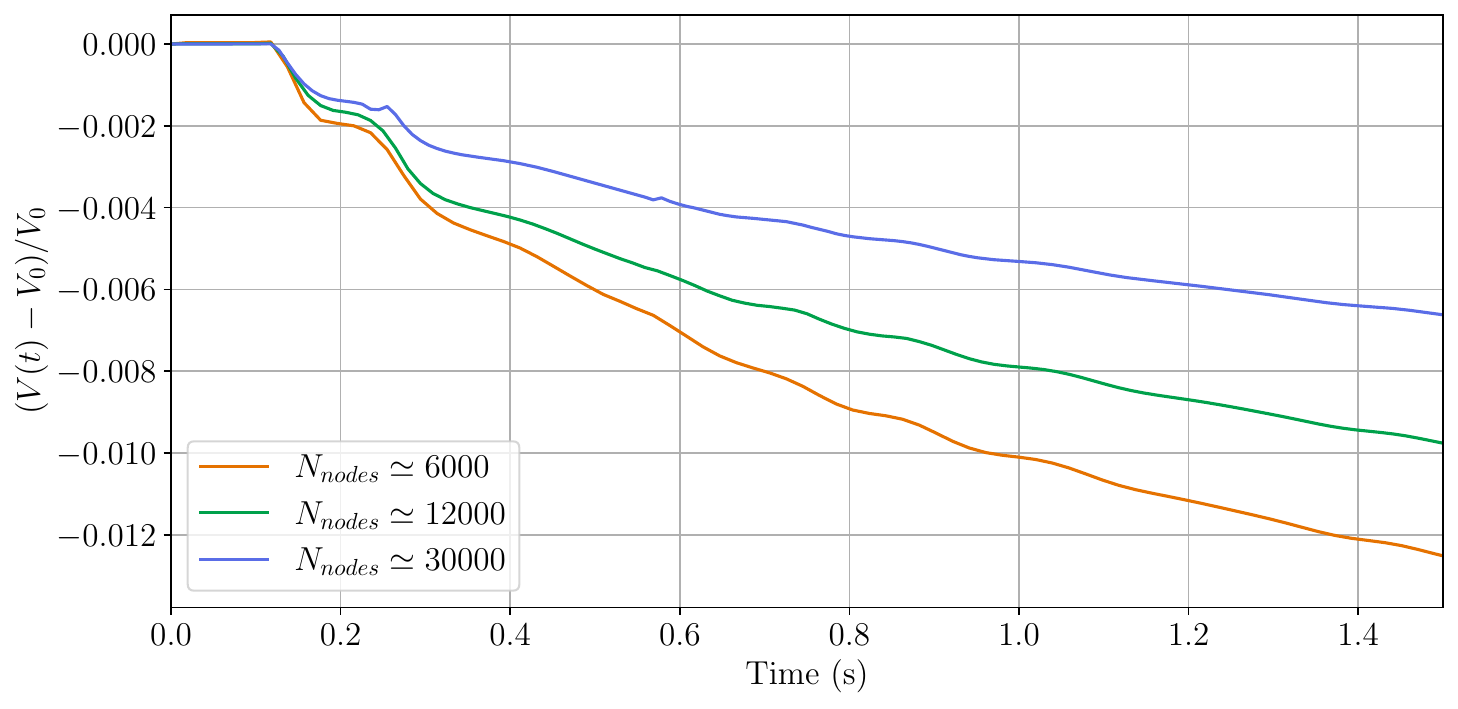}
    \caption{Relative volume change over time for the falling drop case, for different levels of refinement.}\label{fig:drop3D_volume}
\end{figure}

Figure \ref{fig:drop3D_volume} presents the evolution of the relative volume over time for the falling drop case, for different levels of refinement. 
It shows that the winding number approach is indeed able to greatly limit the volume variation. 
For comparison, the standard $\alpha$-shape approach reaches volume increases above $1\%$ as soon as the drop approaches the bulk, and reaches around $8\%$ after $0.25$ $[\mathrm{s}]$.

\subsection{Dam break over an obstacle}
\begin{figure}[H]
    \centering
    \includegraphics[width=\textwidth]{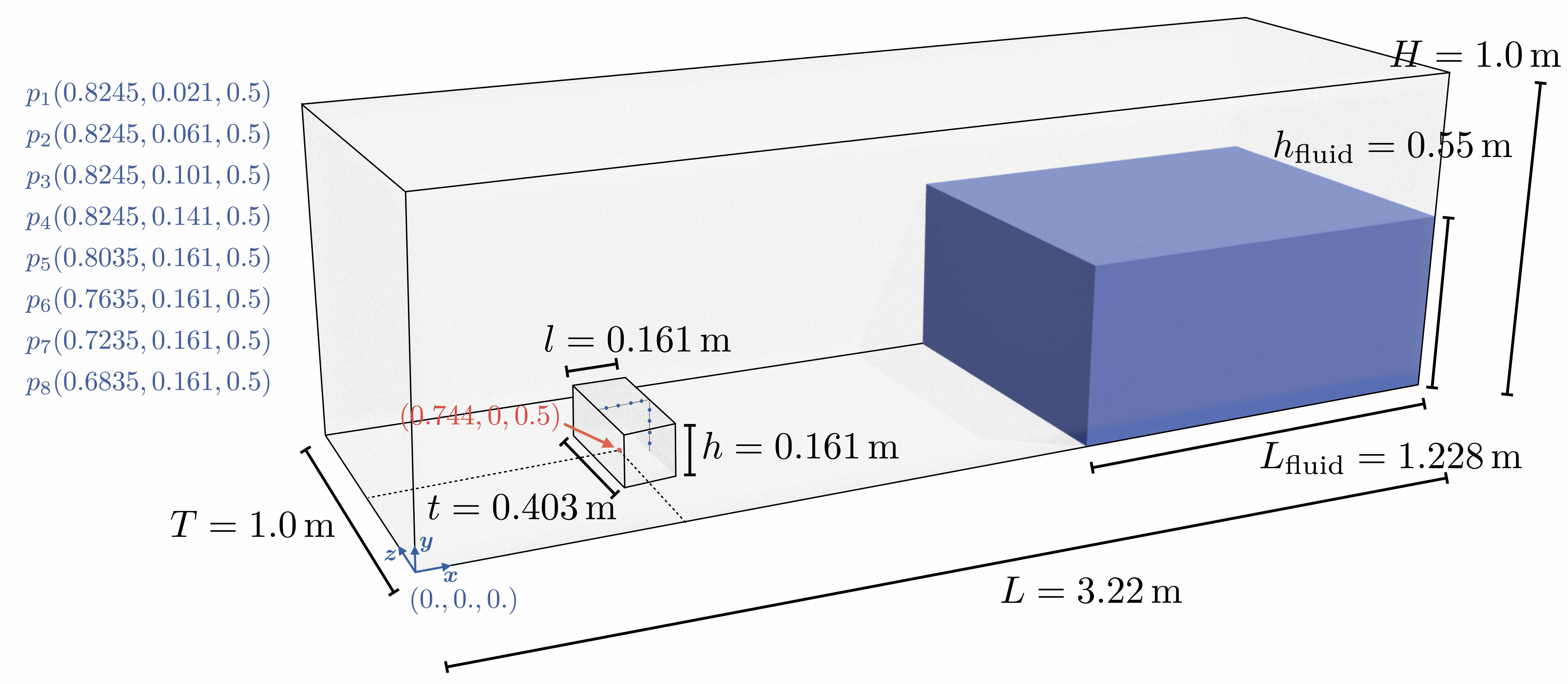}
    \caption{Setup of the dam break experiment.}\label{fig:dam_break_setup}
\end{figure}

\begin{figure}[H]
    \centering
    \includegraphics[width=\textwidth]{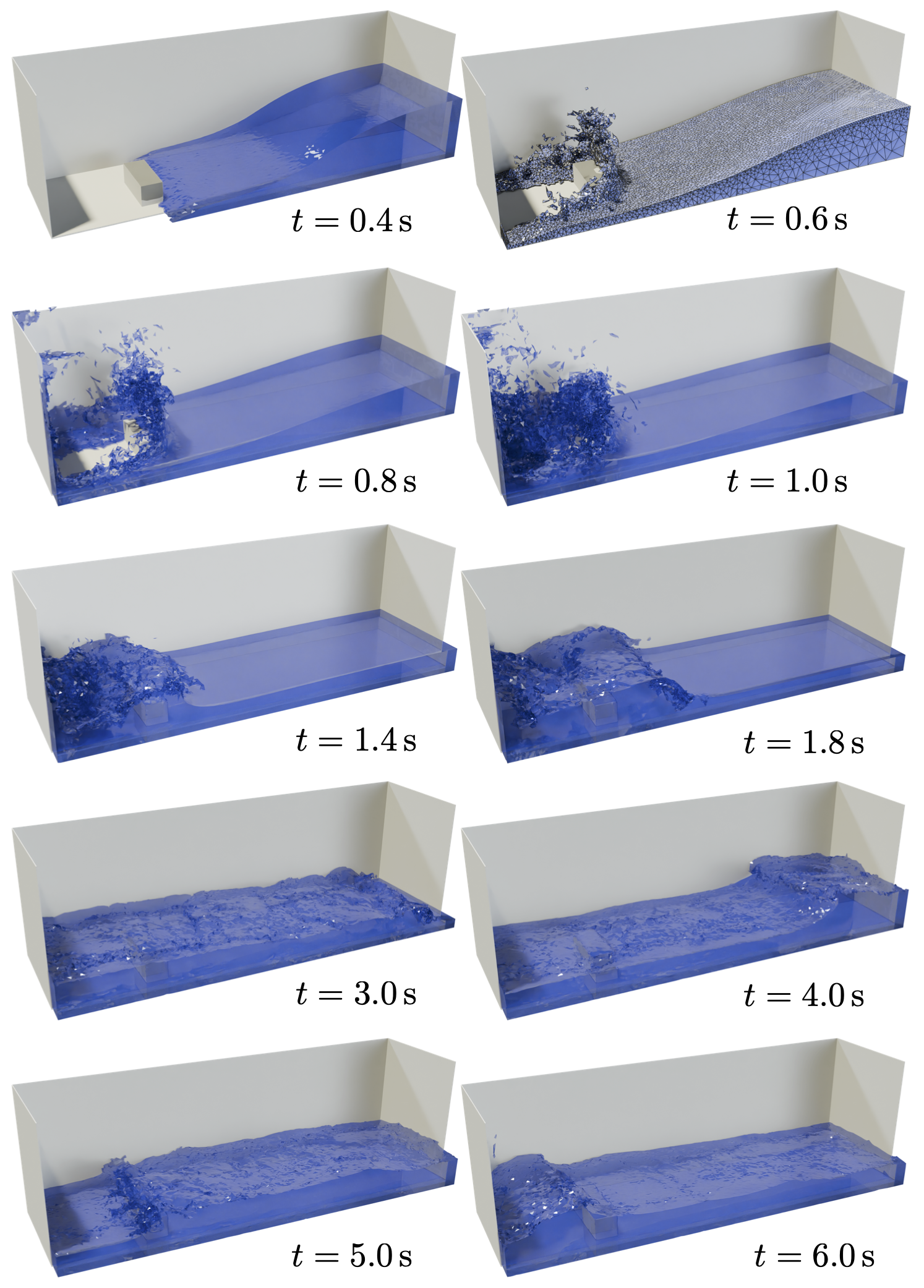}
    \caption{Simulation snapshots of the dam break experiment.}\label{fig:dam_break_snapshots}
\end{figure}

\begin{figure}[H]
    \centering
    \includegraphics[width=.9\textwidth]{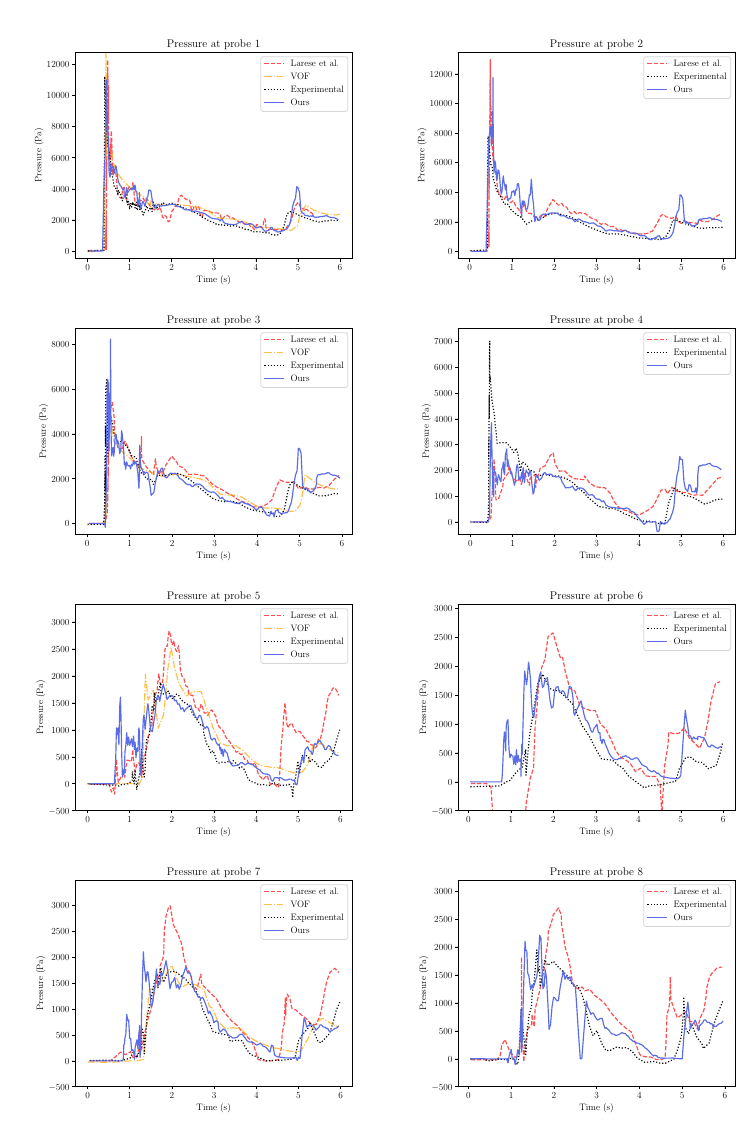}
    \caption{Evolution of the pressure probed at the different locations on the obstacle, comparisons with \cite{larese} (PFEM), \cite{dambreak_exp} (experimental) and \cite{nguyenVOF} (Volume-of-fluid).}\label{fig:dam_break_pressures}
\end{figure}

In this next test case, we analyze the flow of a fluid front hitting an obstacle. 
The initial setup, presented in Figure \ref{fig:dam_break_setup}, consists of a column of water at rest. 
At $t=0$, an imaginary wall is suddenly removed on the left side of the fluid, resulting in the collapse of the column and a fluid front advances towards the left. 
At around $t=0.4$ $[\mathrm{s}]$, the fluid hits a step-like obstacle. 
While some of the water rises over the obstacle, another part flows left and right of it. 
A very complex flow, consisting of many waves hitting the obstacle and the outer walls, develops afterwards. 
In \cite{larese}, the same simulation was performed using the PFEM, and compared to experimental results from \cite{dambreak_exp}. 
For a quantitative comparison, eight pressure probes have been placed on the obstacle to track the evolution of this flow. 
The coordinates of the eight probes are also given in Figure \ref{fig:dam_break_setup}.
A few snapshots of the simulation are presented in Figure \ref{fig:dam_break_snapshots}.
The pressure results are plotted in Figure \ref{fig:dam_break_pressures} and compared to the results of \cite{larese} and \cite{nguyenVOF}, as well as to experimental data from \cite{dambreak_exp}.
The four probes on the vertical wall (probes 1 to 4) show good agreement with the other results, both numerical and experimental. 
The four probes on the top wall of the obstacle (probes 5 to 8) present less accurate results. 
This can mostly be explained by the complexity of the flow at this location, characterized by many splashing events. 

\section{Applications with complex boundaries}\label{sec:applic}
The proposed PFEM is designed to be robust and adaptable to complex geometries. 
The aim of the final two simulations is to show that we are able to perform simulations with complex input geometries.

\subsection{Tire splashing}

\begin{figure}[H]
    \centering
    \includegraphics[width =.7\textwidth]{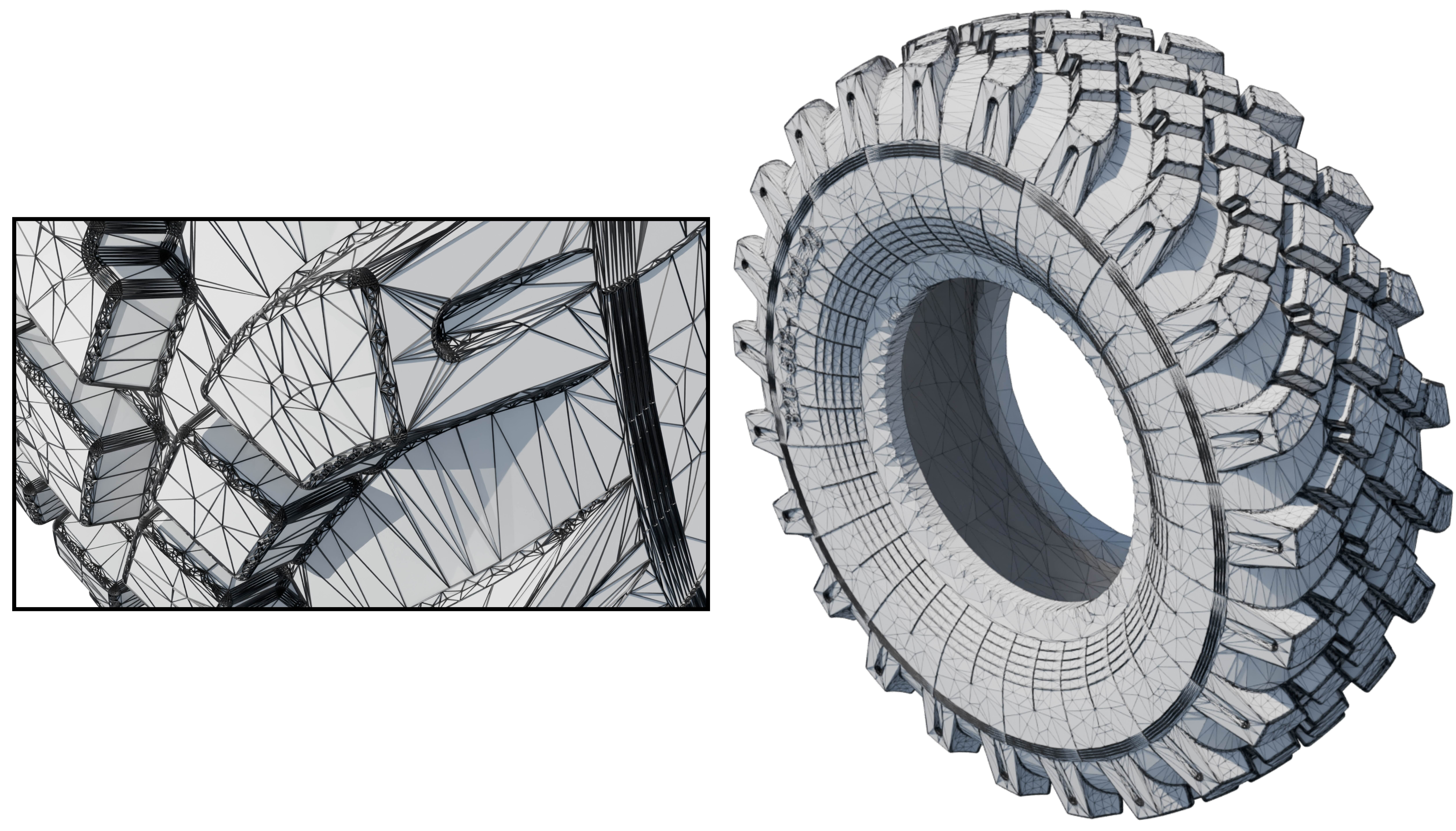}
    \caption{The input mesh of the truck tire.}\label{fig:tire_mesh}
\end{figure}

In this example, we perform a simulation of a tire rolling at a constant speed and crossing a puddle of fluid. 
This generates splashes of the fluid and intricate flow behavior in the tire grooves. 
The input geometry is a truck tire, represented in Figure \ref{fig:tire_mesh}.
The CAD model is extremely complex and contains many poor-quality, elongated elements. 

\begin{figure}[H]
    \centering
    \includegraphics[width = .7\textwidth]{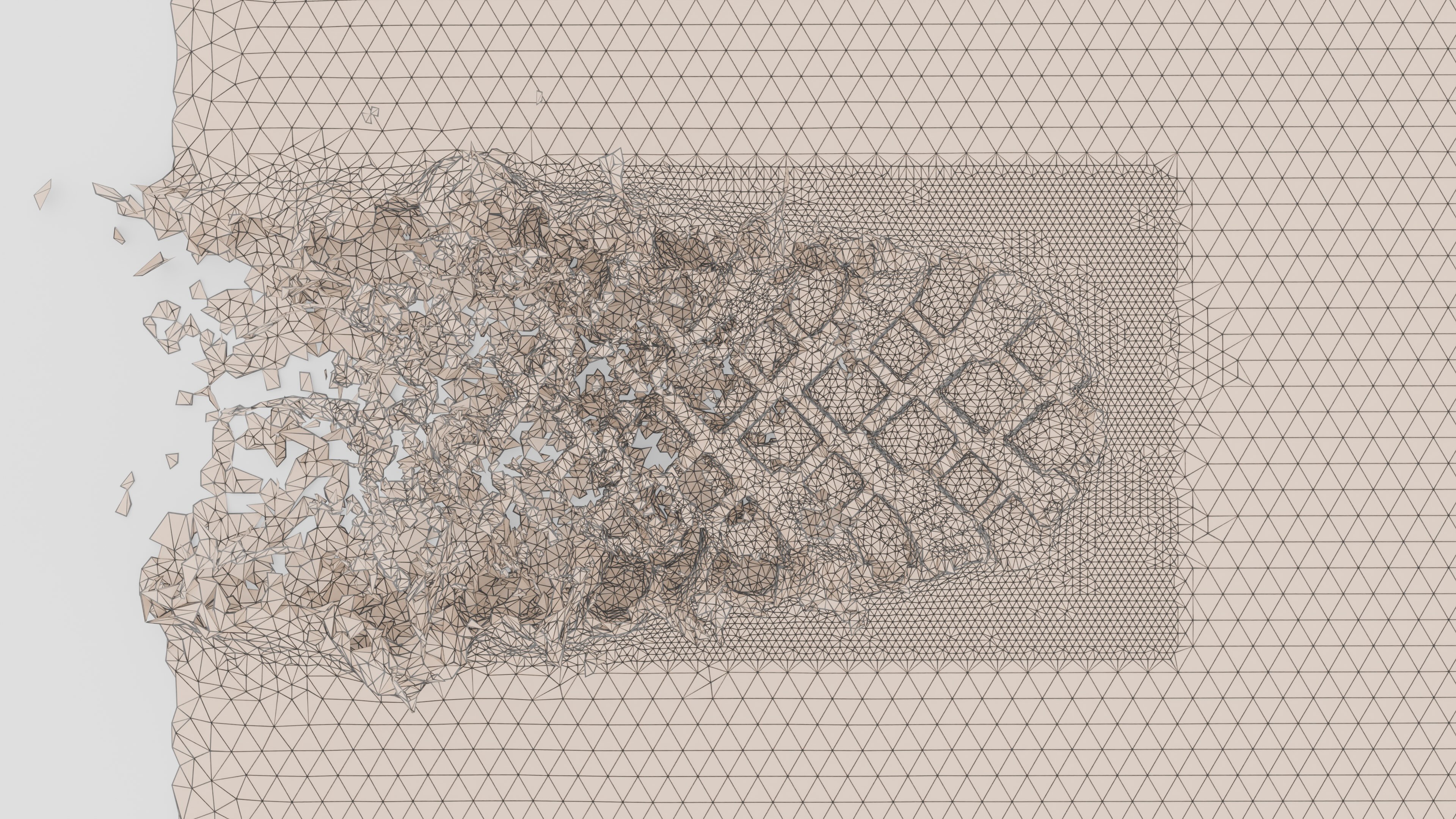}
    \caption{Simulation of a truck tire rolling over mud. Zoom on the mesh refinement and boundary detection.}\label{fig:tire_refinement}
\end{figure}

Using the adaptive refinement approach, we only refine the fluid in the region close to where the fluid is in contact with the tire.
Hence, the fluid mesh is also coarsened after the tire has passed a specific region. 
A snapshot representing the fluid mesh is presented in Figure \ref{fig:tire_refinement}.

\begin{figure}[H]
    \centering
    \includegraphics[width=.9\textwidth]{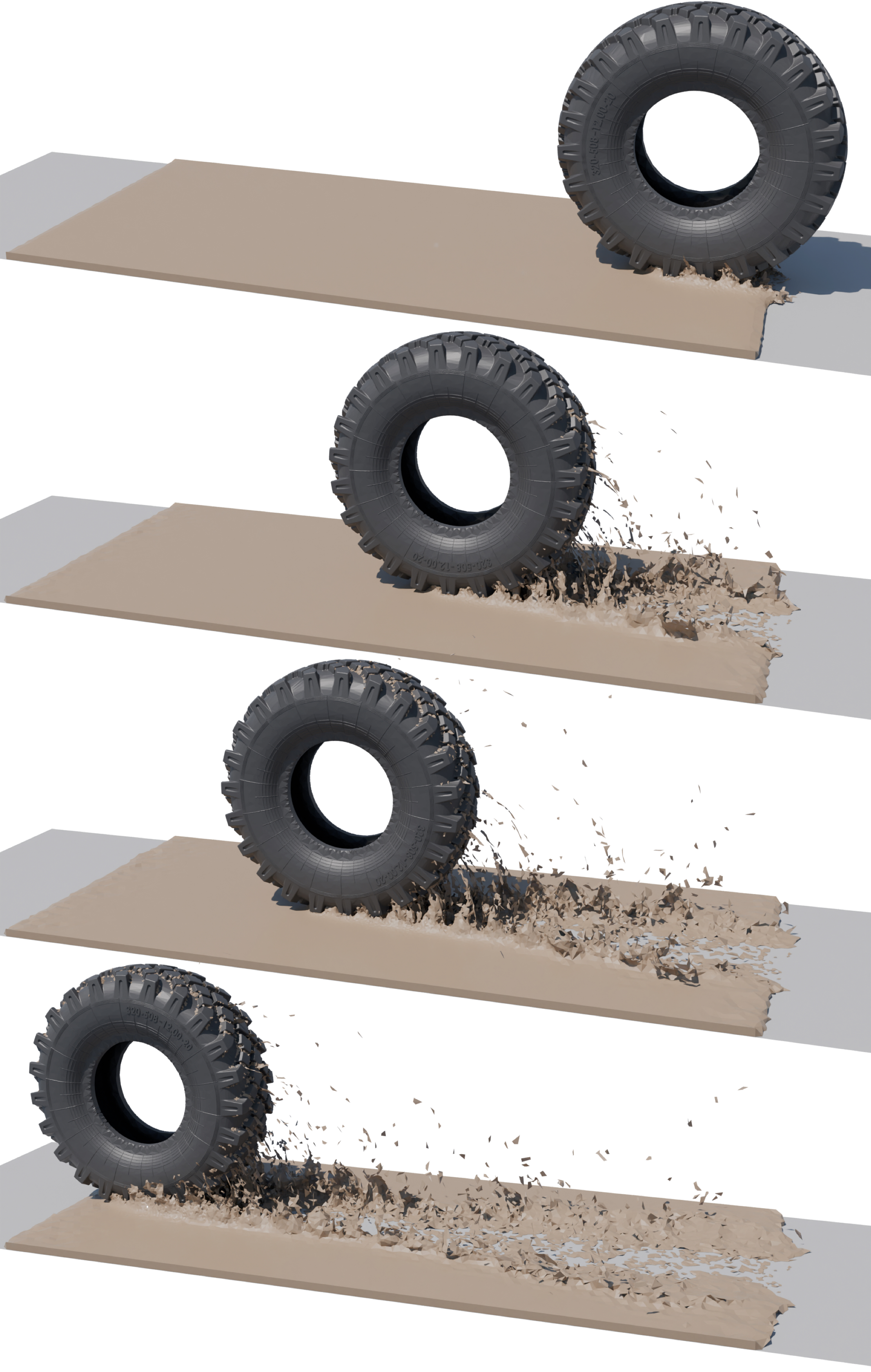}
    \caption{A few snapshots of the tire splashing simulation.}\label{fig:tire_snapshots}
\end{figure}

The truck tire moves at a horizontal speed of 30 $[\mathrm{km\,h^{-1}}]$. 
The fluid has a viscosity of 0.1 $[\mathrm{Pa\,s}]$, and a density of 1000 $[\mathrm{kg\,m^{-3}}]$.
A few snapshots of the simulation are presented in Figure \ref{fig:tire_snapshots}.

\subsection{Simulation on a broken input mesh}
In this final example, we show that the boundary management approach allows us to perform simulations on broken input geometries. 
Non-watertight geometries, as they are often referred to, are not only a hypothetical problem. 
In industry, many assemblies of different parts of a model contain imperfections such as holes and overlaps.  
Fixing these broken designs is very cumbersome.
Automating this is complex and therefore requires a lot of manual effort and time. 

\begin{figure}[H]
    \centering
    \includegraphics[width=.8\textwidth]{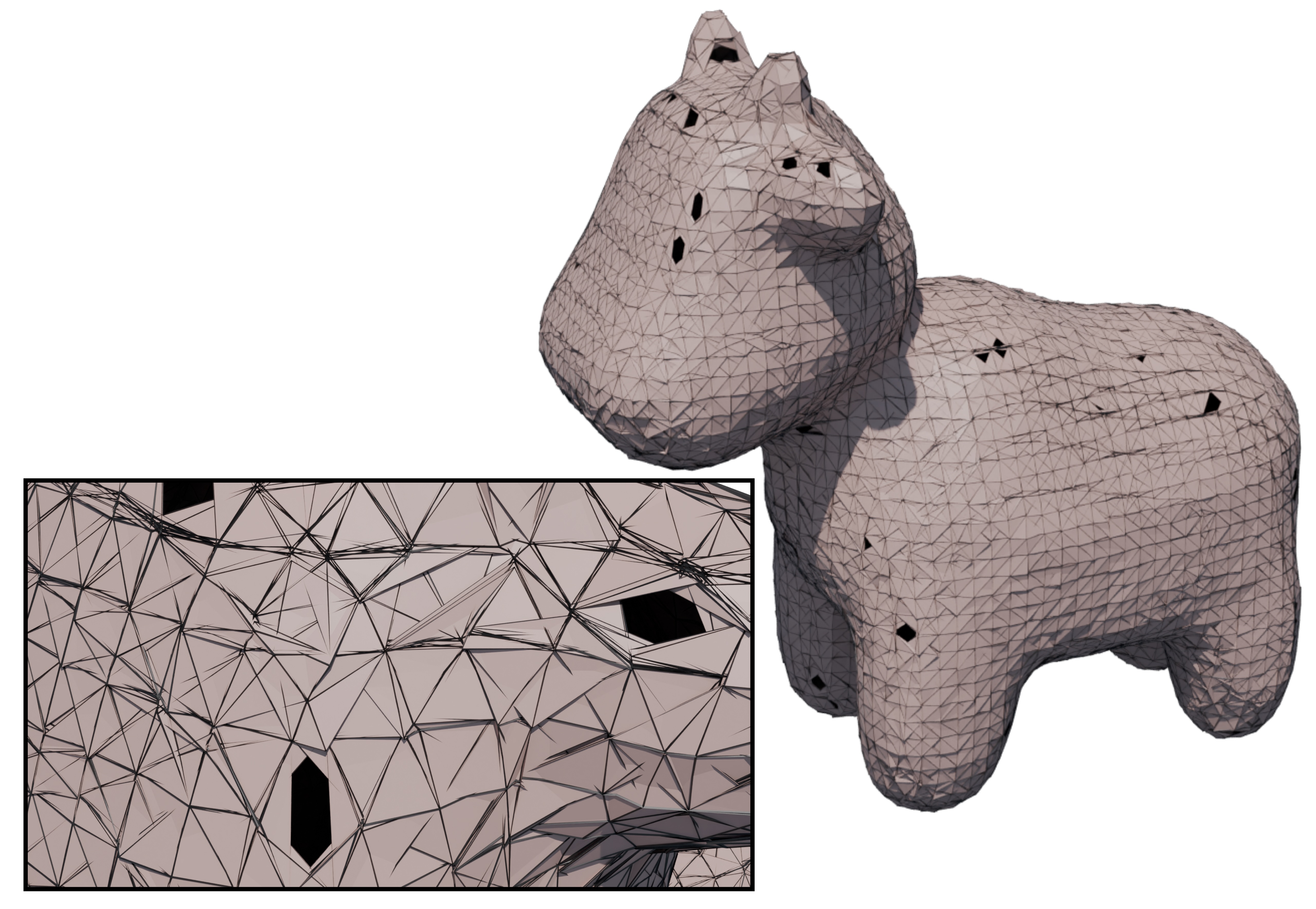}
    \caption{A broken geometry as input, a soup of triangles.}\label{fig:spot_mesh}
\end{figure}

The input may not even be a mesh, but a point cloud. 
An easy way of visualizing this point cloud is by creating a soup of triangles.
The main advantage of these soups of triangles is that they are very easy to construct. 
It can be generated on the point cloud by taking the union of local triangulations about each point formed, for example, by its ten nearest neighbors. 
This results in a highly non-manifold triangulation, but nonetheless gives an idea of the shape of the domain. 
Figure \ref{fig:spot_mesh} presents such a triangle soup.

\begin{figure}[H]
    \centering
    \includegraphics[width=\textwidth]{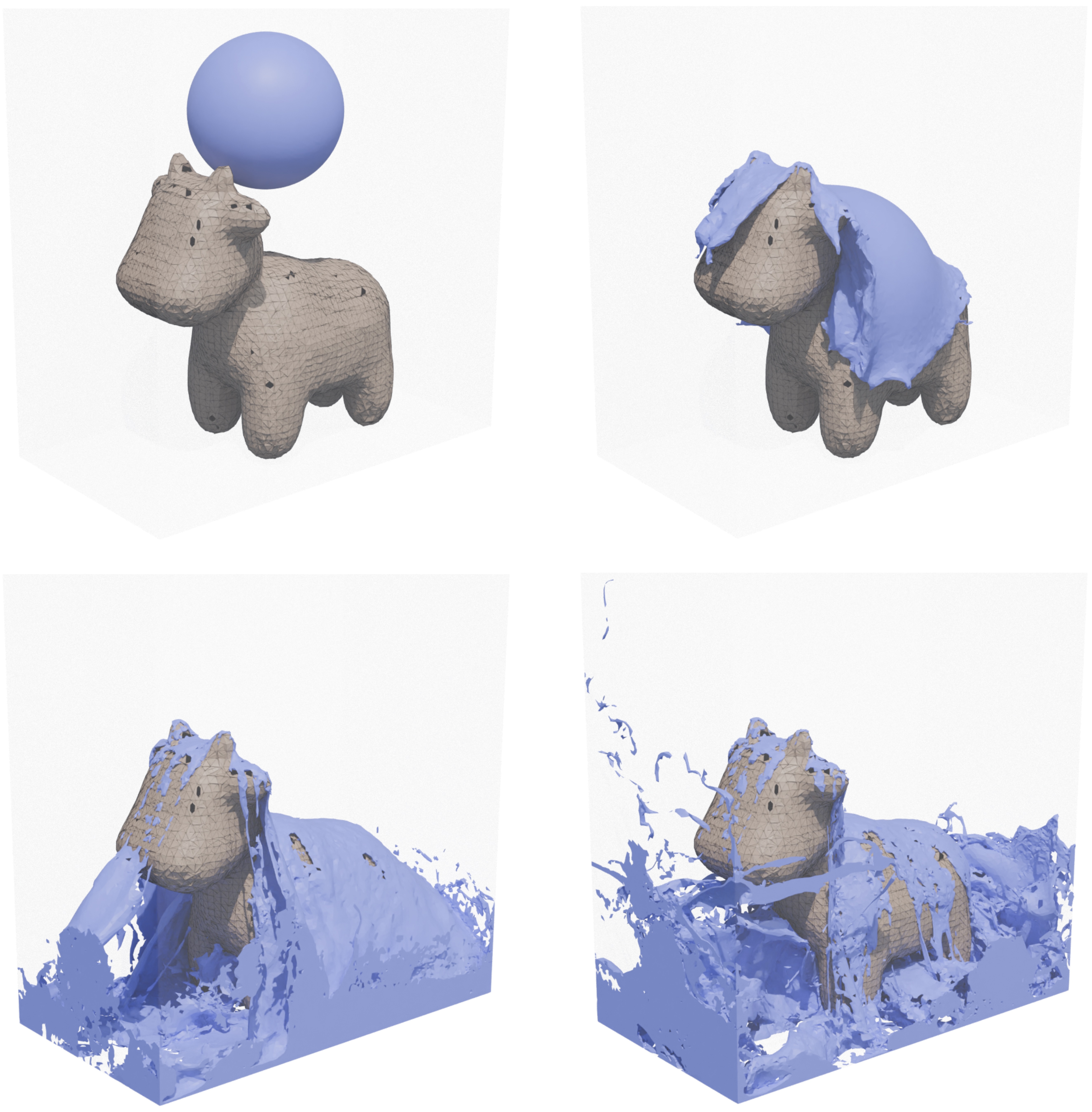}
    \caption{Snapshots of a fluid simulation on a "soup of triangles" geometry.}\label{fig:spot}
\end{figure}

Using our PFEM approach, and specifically the proposed boundary management technique, we can perform simulations around such broken geometries. 
An illustration is presented in Figure \ref{fig:spot}.
As presented before, we simply store all triangles of the boundary in an octree data structure and perform searches during the advection of the particles (step 6 of Figure \ref{fig:pfem_scheme_3D}) to ensure no particles cross the boundary. 
If one does cross a triangle of the boundary, it is reprojected onto the first triangle it crossed.
During the detection of the new fluid domain (step 2 of Figure \ref{fig:pfem_scheme_3D}), any element that is detected as outside of the geometry by the winding number approach is removed from the fluid.
The fluid simulation may present some imperfections where there are large holes in the input geometry, but this must be accepted, as the winding number becomes ambiguous in these regions. 
The main advantage is that the solver remains robust in these complex situations.  

\section{Conclusions}

In this work, we focused on geometric challenges of the particle finite element method.
These are important considerations in the context of highly chaotic three-dimensional free surface flows. 
The mesh is crucial in PFEM, and it plays a much more central role in simulations than in classical finite element methods. 


Three aspects of geometry within the PFEM algorithm have been addressed, forming the basis of the re-meshing procedure.   
The first contribution is to propose a different approach to detect the shape of the domain. 
Using the advected boundary as a predicate to determine which elements are part of the domain, and which are not, allow for more accurate detection of topological changes than with the classical $\alpha$-shape procedure. 
To determine whether an element is inside or outside the domain, the winding number is computed at its center of mass, taking the advected boundary as the closed surface. 
This is a very efficient approach to limiting mass conservation issues when topological changes occur. 

The second contribution is a mesh adaptation procedure. 
The process is composed of two steps. 
First, a surface refinement performs particle insertion through an edge-splitting approach. 
This maintains the shape of the fluid, and the quality of the boundary elements remains sufficiently high. 
Second, the bulk is refined by inserting new particles at element circumcenters, based on the user-defined size field. 
This two-phase approach ensures that the shape of the domain is maintained and allows for local mesh refinement and adaptation. 

The final contribution presents an approach to manage boundaries in complex domains by decoupling the fluid domain from the solid geometry. 
With the use of three simple geometric queries -- inside/outside detection, triangle-segment intersections, and point projections -- boundaries can be easily coloured. 
This results in a robust and efficient method to manage interactions of fluids with solid walls. 
Moreover, slip boundary conditions are easily applied with this approach. 

The various implementations have been illustrated through different applications, including quantitative comparisons of a dam break flow over an obstacle and visual results of a truck tire rolling over a fluid puddle and generating splashing phenomena.
The final simulation shows that the approach is robust even with poor-quality input meshes such as soups of triangles.
These applications show that the proposed method is both quantitatively accurate and can be applied in the presence of complex, possibly non-watertight geometries.

\begin{figure}[H]
    \centering
    \includegraphics[width=.8\textwidth]{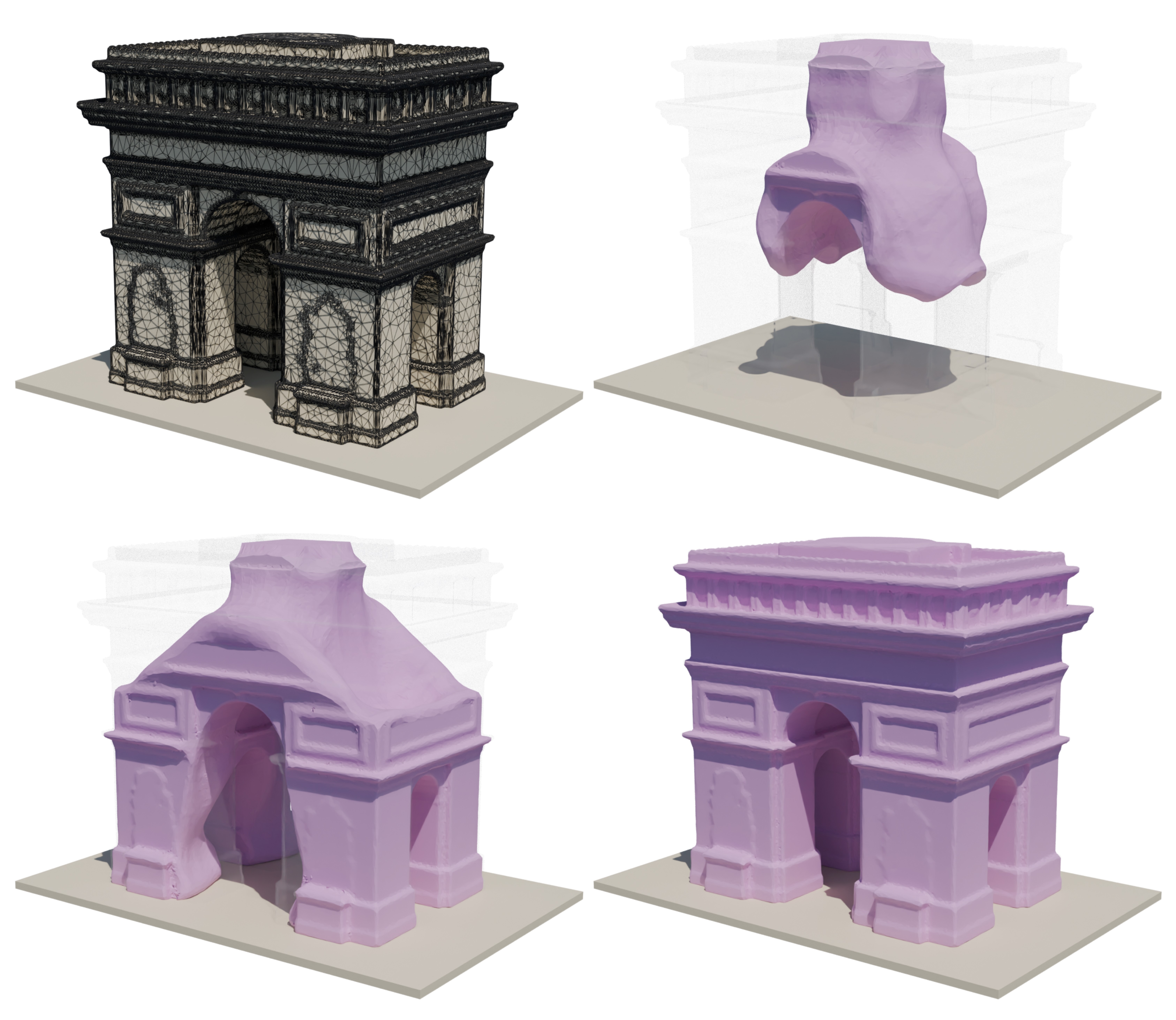}
    \caption{A complex model is filled to generate a simulation-ready mesh.}\label{fig:triomphe}
\end{figure}

Future developments include efficiency improvements by addressing parallelization and a GPU-based implementation. 
The the presence of slivers inside meshes and their effect on the quality of results is also a subject of future analysis.  
Finally, Figure \ref{fig:triomphe} presents another perspective of PFEM simulations, namely, mesh generation on broken inputs.

\section{Acknowledgments}
Thomas Leyssens is a Research Fellow of the F.R.S.-FNRS.
Computational resources have been provided by the supercomputing facilities of the Université catholique de Louvain (CISM/UCL) and the Consortium des Équipements de Calcul Intensif en Fédération Wallonie Bruxelles (CÉCI) funded by the Fond de la Recherche Scientifique de Belgique (F.R.S.-FNRS) under convention 2.5020.11 and by the Walloon Region.
This project has received funding from the European Research Council (ERC) under the European Union's Horizon research and innovation program (Grant agreement no 101 071 255).

\printbibliography

@article{celestin,
  title={One machine, one minute, three billion tetrahedra},
  author={Marot, C{\'e}lestin and Pellerin, Jeanne and Remacle, Jean-Fran{\c{c}}ois},
  journal={International Journal for Numerical Methods in Engineering},
  volume={117},
  number={9},
  pages={967--990},
  year={2019},
  publisher={Wiley Online Library}
}

@article{franciMass,
  title={On the effect of standard PFEM remeshing on volume conservation in free-surface fluid flow problems},
  author={Franci, Alessandro and Cremonesi, Massimiliano},
  journal={Computational particle mechanics},
  volume={4},
  number={3},
  pages={331--343},
  year={2017},
  publisher={Springer}
}

@inproceedings{chew,
  title={Guaranteed-quality mesh generation for curved surfaces},
  author={Chew, L Paul},
  booktitle={Proceedings of the ninth annual symposium on Computational geometry},
  pages={274--280},
  year={1993}
}

@article{mpm,
  title={Material point method after 25 years: Theory, implementation, and applications},
  author={De Vaucorbeil, Alban and Nguyen, Vinh Phu and Sinaie, Sina and Wu, Jian Ying},
  journal={Advances in applied mechanics},
  volume={53},
  pages={185--398},
  year={2020},
  publisher={Elsevier}
}

@article{falla,
  title={Mesh adaption for two-dimensional bounded and free-surface flows with the particle finite element method},
  author={Falla, Romain and Bobach, Billy-Joe and Boman, Romain and Ponthot, Jean-Philippe and Terrapon, Vincent E},
  journal={Computational Particle Mechanics},
  pages={1--28},
  year={2023},
  publisher={Springer}
}

@article{cerquagliaFSI,
  title={A fully partitioned Lagrangian framework for FSI problems characterized by free surfaces, large solid deformations and displacements, and strong added-mass effects},
  author={Cerquaglia, Marco-Lucio and Thomas, David and Boman, Romain and Terrapon, Vincent and Ponthot, J-P},
  journal={Computer Methods in Applied Mechanics and Engineering},
  volume={348},
  pages={409--442},
  year={2019},
  publisher={Elsevier}
}

@article{gmsh,
  title={Gmsh: A 3-D finite element mesh generator with built-in pre-and post-processing facilities},
  author={Geuzaine, Christophe and Remacle, Jean-Fran{\c{c}}ois},
  journal={International journal for numerical methods in engineering},
  volume={79},
  number={11},
  pages={1309--1331},
  year={2009},
  publisher={Wiley Online Library}
}

@article{migflow,
  title={Implementation of an unresolved stabilised FEM--DEM model to solve immersed granular flows},
  author={Constant, Matthieu and Dubois, Fr{\'e}d{\'e}ric and Lambrechts, Jonathan and Legat, Vincent},
  journal={Computational Particle Mechanics},
  volume={6},
  pages={213--226},
  year={2019},
  publisher={Springer}
}

@article{pspg,
  title={A new finite element formulation for computational fluid dynamics: V. Circumventing the Babu{\v{s}}ka-Brezzi condition: A stable Petrov-Galerkin formulation of the Stokes problem accommodating equal-order interpolations},
  author={Hughes, Thomas JR and Franca, Leopoldo P and Balestra, Marc},
  journal={Computer Methods in Applied Mechanics and Engineering},
  volume={59},
  number={1},
  pages={85--99},
  year={1986},
  publisher={Elsevier}
}

@inproceedings{octree,
  title={Efficient synthetic image generation of arbitrary 3-D objects},
  author={Meagher, Donald},
  booktitle={Proceeding of the IEEE Conference on Pattern Recognition and Image Processing},
  volume={473},
  year={1982}
}

@article{leyssens,
  title={A Delaunay refinement algorithm for the particle finite element method applied to free surface flows},
  author={Leyssens, Thomas and Henry, Michel and Lambrechts, Jonathan and Remacle, Jean-Fran{\c{c}}ois},
  journal={International Journal for Numerical Methods in Engineering},
  pages={e7554},
  publisher={Wiley Online Library},
  year={2024}
}

@article{vajont,
  title={3D simulation of Vajont disaster. Part 1: Numerical formulation and validation},
  author={Franci, Alessandro and Cremonesi, Massimiliano and Perego, Umberto and Crosta, Giovanni and O{\~n}ate, Eugenio},
  journal={Engineering Geology},
  volume={279},
  pages={105854},
  year={2020},
  publisher={Elsevier}
}

@article{alphaShapes3D,
  title={Three-dimensional alpha shapes},
  author={Edelsbrunner, Herbert and M{\"u}cke, Ernst P},
  journal={ACM Transactions On Graphics (TOG)},
  volume={13},
  number={1},
  pages={43--72},
  year={1994},
  publisher={ACM New York, NY, USA}
}

@article{rivara84,
  title={Algorithms for refining triangular grids suitable for adaptive and multigrid techniques},
  author={Rivara, M Cecilia},
  journal={International journal for numerical methods in Engineering},
  volume={20},
  number={4},
  pages={745--756},
  year={1984},
  publisher={Wiley Online Library}
}

@article{rivara97,
  title={New longest-edge algorithms for the refinement and/or improvement of unstructured triangulations},
  author={Rivara, Maria-Cecilia},
  journal={International journal for numerical methods in Engineering},
  volume={40},
  number={18},
  pages={3313--3324},
  year={1997},
  publisher={Wiley Online Library}
}

@article{rosenberg_proof_bisect,
  title={A lower bound on the angles of triangles constructed by bisecting the longest side},
  author={Rosenberg, Ivo G and Stenger, Frank},
  journal={Mathematics of Computation},
  volume={29},
  number={130},
  pages={390--395},
  year={1975}
}

@article{halfEdge,
  title={Finding the intersection of two convex polyhedra},
  author={Muller, David E. and Preparata, Franco P.},
  journal={Theoretical Computer Science},
  volume={7},
  number={2},
  pages={217--236},
  year={1978},
  publisher={Elsevier}
}

@article{vortex3D,
  title={High-resolution conservative algorithms for advection in incompressible flow},
  author={Leveque, Randall J},
  journal={SIAM Journal on Numerical Analysis},
  volume={33},
  number={2},
  pages={627--665},
  year={1996},
  publisher={SIAM}
}

@article{meduri2019,
  title={An efficient runtime mesh smoothing technique for 3D explicit Lagrangian free-surface fluid flow simulations},
  author={Meduri, S and Cremonesi, M and Perego, U},
  journal={International Journal for Numerical Methods in Engineering},
  volume={117},
  number={4},
  pages={430--452},
  year={2019},
  publisher={Wiley Online Library}
}

@article{windingNumber2,
  title={Fast winding numbers for soups and clouds},
  author={Barill, Gavin and Dickson, Neil G and Schmidt, Ryan and Levin, David IW and Jacobson, Alec},
  journal={ACM Transactions on Graphics (TOG)},
  volume={37},
  number={4},
  pages={1--12},
  year={2018},
  publisher={ACM New York, NY, USA}
}

@article{windingNumber1,
  title={Robust inside-outside segmentation using generalized winding numbers},
  author={Jacobson, Alec and Kavan, Ladislav and Sorkine-Hornung, Olga},
  journal={ACM Transactions on Graphics (TOG)},
  volume={32},
  number={4},
  pages={1--12},
  year={2013},
  publisher={ACM New York, NY, USA}
}

@article{henry2025multiscale,
  title={Multiscale FEM-DEM model for spontaneous droplet digging in a hot granular bed},
  author={Henry, Michel and Coppin, Nathan and Dorbolo, St{\'e}phane and Legat, Vincent and Lambrechts, Jonathan},
  journal={International Journal of Heat and Mass Transfer},
  volume={241},
  pages={126755},
  year={2025},
  publisher={Elsevier}
}

@article{LEYSSENS2025114082,
title = {A coupled PFEM-DEM model for fluid-granular flows with free surface dynamics applied to landslides},
journal = {Journal of Computational Physics},
volume = {537},
pages = {114082},
year = {2025},
issn = {0021-9991},
doi = {https://doi.org/10.1016/j.jcp.2025.114082},
url = {https://www.sciencedirect.com/science/article/pii/S0021999125003651},
author = {Thomas Leyssens and Michel Henry and Jonathan Lambrechts and Vincent Legat and Jean-François Remacle},
}

@article{rizzieri,
  title={Numerical simulation of the extrusion and layer deposition processes in 3D concrete printing with the Particle Finite Element Method},
  author={Rizzieri, Giacomo and Ferrara, Liberato and Cremonesi, Massimiliano},
  journal={Computational Mechanics},
  volume={73},
  number={2},
  pages={277--295},
  year={2024},
  publisher={Springer}
}

@article{fevrier,
  title={Simulation of melt pool dynamics including vaporization using the particle finite element method},
  author={F{\'e}vrier, Simon and Fern{\'a}ndez, Eduardo and Lacroix, Martin and Boman, Romain and Ponthot, Jean-Philippe},
  journal={Computational Mechanics},
  pages={1--29},
  year={2024},
  publisher={Springer}
}

@article{pic,
  title={FLIP: a low-dissipation, particle-in-cell method for fluid flow},
  author={Brackbill, Jeremiah U and Kothe, Douglas B and Ruppel, Hans M},
  journal={Computer Physics Communications},
  volume={48},
  number={1},
  pages={25--38},
  year={1988},
  publisher={Elsevier}
}

@article{tetgen,
  title={TetGen, a Delaunay-based quality tetrahedral mesh generator},
  author={Hang, Si},
  journal={ACM Trans. Math. Softw},
  volume={41},
  number={2},
  pages={11},
  year={2015}
}

@phdthesis{marotThese,
  title={Parallel tetrahedral mesh generation},
  author={Marot, C{\'e}lestin},
  year={2020},
  school={Ph. D. dissertation, UCL-Universit{\'e} Catholique de Louvain, Belgium, 2020~…}
}

@article{imaging_alphashape,
  title={A Precisely One-Step Registration Methodology for Optical Imagery and LiDAR Data Using Virtual Point Primitives},
  author={Yao, Chunjing and Ma, Hongchao and Luo, Wenjun and Ma, Haichi},
  journal={Remote Sensing},
  volume={13},
  number={23},
  pages={4836},
  year={2021},
  publisher={MDPI}
}

@incollection{alphashapeSurvey,
  title={Alpha shapes-a survey},
  author={Edelsbrunner, Herbert},
  booktitle={Tessellations in the sciences: Virtues, techniques and applications of geometric tilings},
  year={2011}
}

@article{larese,
  title={Validation of the particle finite element method (PFEM) for simulation of free surface flows},
  author={Larese, ANTONIA and Rossi, Riccardo and O{\~n}ate, E and Idelsohn, SR},
  journal={Engineering Computations},
  volume={25},
  number={4},
  pages={385--425},
  year={2008},
  publisher={Emerald Group Publishing Limited}
}

@article{dambreak_exp,
  title={Test-case 2, 3D dambreaking},
  author={Issa, Reza and Violeau, Damien},
  journal={ERCOFTAC, SPH European Research Interest Community SIG},
  year={2006}
}

@inproceedings{schroeder1992decimation,
  title={Decimation of triangle meshes},
  author={Schroeder, William J and Zarge, Jonathan A and Lorensen, William E},
  booktitle={Proceedings of the 19th annual conference on Computer graphics and interactive techniques},
  pages={65--70},
  year={1992}
}

@article{nguyenVOF,
  title={A free surface flow solver for complex three-dimensional water impact problems based on the VOF method},
  author={Nguyen, Van-Tu and Park, Warn-Gyu},
  journal={International Journal for Numerical Methods in Fluids},
  volume={82},
  number={1},
  pages={3--34},
  year={2016},
  publisher={Wiley Online Library}
}

@article{PVEM,
  title={Particle Virtual Element Method (PVEM): an agglomeration technique for mesh optimization in explicit Lagrangian free-surface fluid modelling},
  author={Fu, Cheng and Cremonesi, Massimiliano and Perego, Umberto and Hudobivnik, Bla{\v{z}} and Wriggers, Peter},
  journal={Computer Methods in Applied Mechanics and Engineering},
  volume={433},
  pages={117461},
  year={2025},
  publisher={Elsevier}
}

@article{kuvcera2016necessary,
  title={On necessary and sufficient conditions for finite element convergence},
  author={Ku{\v{c}}era, Vaclav},
  journal={arXiv preprint arXiv:1601.02942},
  year={2016}
}

@article{hannukainen2012maximum,
  title={The maximum angle condition is not necessary for convergence of the finite element method},
  author={Hannukainen, Antti and Korotov, Sergey and Krizek, Michal},
  journal={Numerische mathematik},
  volume={120},
  number={1},
  pages={79--88},
  year={2012},
  publisher={Springer}
}

@article{quiriny2024tempered,
  title={The tempered finite element method},
  author={Quiriny, Antoine and Kucera, Vaclav and Lambrechts, Jonathan and Mo{\"e}s, Nicolas and Remacle, Jean-Fran{\c{c}}ois},
  journal={arXiv preprint arXiv:2411.17564},
  year={2024}
}

@article{marot2019one,
  title={One machine, one minute, three billion tetrahedra},
  author={Marot, C{\'e}lestin and Pellerin, Jeanne and Remacle, Jean-Fran{\c{c}}ois},
  journal={International Journal for Numerical Methods in Engineering},
  volume={117},
  number={9},
  pages={967--990},
  year={2019},
  publisher={Wiley Online Library}
}
\end{document}